\documentclass{aa} 
\usepackage{color,graphicx}
\usepackage{soul}       
\usepackage{ulem}   
\usepackage{lineno}
\renewcommand{\theenumii}{\theenumi.\arabic{enumii}.}   
\definecolor{cyan}{rgb}{0.23,0.90,0.90}

\definecolor{magenta}{rgb}{0.93,0.20,0.90}

\definecolor{ForestGreen}{rgb}{0.13,0.80,0.40}

\definecolor{red}{rgb}{0.93,0.20,0.30}

\newcommand{\bo}[1]{\it{#1}}
\begin{document} 
   \title{Search for shower's duplicates at the IAU MDC. }
   \subtitle{Methods and general results. }
   \author{T.J. Jopek\inst{1}
     \and
     L. Neslu\v{s}an\inst{2}
      \and
     R. Rudawska\inst{3, 4}
     \and
     M. Hajdukov\'{a} \inst{5}
     }
   \institute{
            {Astronomical Observatory Institute, Faculty of Physics, A.M. University, Pozna\'{n}, Poland,
             \email{jopek@amu.edu.pl}
            \and
            Astronomical Institute, Slovak Academy of Sciences, Tatranska              Lomnica, Slovakia,
            \email{ne@ta3.sk}}
            \and
            RHEA Group, Noordwijk, The Netherlands,
            \and
            ESA ESTEC, Noordwijk, The Netherlands,
            \email{Regina.Rudawska@ ext.esa.int}
            \and
            Astronomical Institute, Slovak Academy of Sciences, Bratislava, Slovakia,
            \email{Maria.Hajdukova@savba.sk}
             }
   \date{Received \today; accepted  202?-??}
%
  \abstract
  {The meteor shower database of the IAU Meteor Data Center (MDC) is used by the whole community of meteor astronomers. Observers submit both new and known meteor shower parameters to the MDC. It may happen that a new observation of an already known meteor shower is submitted as a discovery of a new shower. Then, a duplicate shower appears in the MDC.  
   On the other hand, the observers may provide  data which, in their opinion, is another set of parameters of an already existing shower. However, if this is not true, we can talk about a shower that is a false-duplicate of a known meteor shower. \\
   The MDC database contains such duplicates and false-duplicates, so it is desirable to detect them among the streams already in the database and those delivered to the database as new streams. }
   {We aim to develop a method for objective detection of duplicates among meteor showers and apply it to the MDC. The method will also enable us to verify whether various sets of parameters of the same shower are compatible and, thus, reveal the false-duplicates.}
   {We suggest two methods based on cluster analyses and two similarity functions  among geocentric and heliocentric shower parameters collected in the MDC. 
  }
   {A number of results of varying significance were obtained. Seven new showers represented by two or more parameter sets were discovered. 30 times there was full agreement between our results and those reported in the MDC database. 20 times the same duplicates as given in the MDC, were found only by one method.  We found 34 multi-solution showers for which the number of the same duplicates found by both method is close to the corresponding number in the MDC database.  However for 56 multi-solution showers listed in the MDC no duplicates were found by any of the applied methods. 
  }
   {The obtained results confirmed the effectiveness of the proposed approach of identifying duplicates. We have shown that in order to detect and verify duplicate meteor showers, it is possible to apply the objective proposal instead of the subjective approach used so far. We consider the identification of 87 problematic cases in the MDC database, among which at least some duplicates were misclassified, to be a particularly important result. 
The correction of these cases will significantly improve the content of the MDC database.}
   \keywords{catalogs -- methods: data analysis -- meteorites, meteors, meteoroids}
\maketitle
%
\section{Introduction}
\label{Intro}
In the case of asteroids (comets), it sometimes happens that observations of a new object, after some time, are linked to its observations made in the past. Thus, the 'new' object turned out to be an object already known.
In meteor astronomy, we do not know this kind of cases. In fact, meteoroids falling into the Earth's atmosphere are crushed and in the form of dust, and sometimes larger fragments fall onto the Earth's surface.\footnote{
However, there are exceptions --- the occurrence of an Earth-grazing fireball, a very bright meteoroid that enters Earth’s atmosphere and leaves again, see e.g. \citet{1991JIMO...19...13S}.
}

But in case of the meteoroid streams, we are dealing with an analogy called --- duplicates.  
In particular, we intend to deal with the duplicate showers in the IAU MDC database. This database also includes showers whose average parameters were determined by two or more author teams. Hereafter, we refer to each set of the mean parameters, published by one author team, as a ``solution''.  Hence, in the remainder of this article, if we have more than one solution for a given shower in the MDC database, we will
will use the term multi-solution shower (MSS) and we refer to each of such solutions as "duplicates". Otherwise, we will refer to a single-solution shower (SSS).

When the MDC list was created, no general criteria were adopted to distinguish whether a newly found solution is another solution of already existing shower or if it is the first solution of new shower; the individual author teams proceeded according to their own approaches, but evidently used different criteria. 

However, an MSS solution might have been incorrectly assigned in the MDC and is a stand-alone shower, i.e. not a duplicate of the shower in question. Such a solution is called a ``false duplicate''.

The duplicate showers we are searching for (the redundant showers in the MDC) are showers which were submitted as new, autonomous showers to the MDC, but, actually, are other solutions of previously submitted showers. To each such solution, it is also referred to as a ``duplicate''. Hence, a duplicate can be  a redundant old (already existing in the database) or new (just supplied to the database) solution of a given shower. And it is something obvious that a correct list of showers in the MDC should not contain MSS that consist of false duplicates.

If an autonomous shower is found to be a duplicate of another shower, its solution will be appended to that shower and the autonomous shower will be moved to the List of Removed Showers. Of course, it is possible that some MSS solutions will be found to be duplicates of other showers and the rest of the MSS solutions will remain as duplicates of the original showers. In that case, the original shower will be retained, but with fewer solutions.

Within the MDC shower database, one can now find solutions that have been both correctly and incorrectly classified. To provide a clearer understanding of the terms used, we present a summary of them. \\
The correctly classified solutions are:\\
{\it (i) an autonomous solution of an SSS},\\
{\it (ii) a duplicate solution within an MSS} (or, simply, a duplicate).\\
The solutions that were classified incorrectly and need to be re-classified are:\\
{\it (iii) a false duplicate solution} --- a solution incorrectly classified as a duplicate solution of an MSS; after correction, it will be re-classified as a new SSS or as another solution of a different MSS, \\
{\it (iv) a false autonomous solution} --- after correction, it will be re-classified as a duplicate solution of a different MSS.\\

In order to determine whether we are dealing with duplicate meteor showers, the methods developed to identify comets and asteroids observed in successive apparitions are proving ineffective.  The main obstacle are the single-apparition nature of the meteor phenomenon and the markedly lower precision of its observation.

Moreover, the geocentric and heliocentric parameters of the repeatedly observed meteoroid stream can differ significantly. The different observational techniques and weather conditions can cause that independent observers detect the meteors of different size distributions and/or in the different periods of the shower's activity. As result, the new and old mean characteristics of a given shower can be quite different.

Hence, the question whether we are dealing with a duplicate  of a shower or not is not trivial. 
One can find several publications mentioning the occurrence of similar showers (shower's duplicates in our terminology) in the MDC database, e.g.: \citet[][]{2012JIMO...40..166H, 2014pim4.conf..126A, 2017eMetN...2...27K, 2020eMetN...5...93K}. Usually, authors provide a list of pairs that they consider to be similar showers, and recommend removing the shower in question from the database and introducing it as another solution to an already known shower, \citep[e.g.][]{2012JIMO...40..166H}. 

To date, however, the identification of a duplicate meteor shower and the associated recommendation have been based on more or less subjective considerations.  
Therefore, our work has focused on developing of an objective method for detecting duplicates among  meteor showers in the MDC database. Two approaches were used: the method  based on orbital similarity and the method developed by us called "maximum-$\sigma$'s" criterion based on direct comparison of selected geocentric and heliocentric shower parameters.

%
\section{Meteor shower data used}
\label{MDC-data}
As to December 2022, the MDC database contained data of $920$ meteor showers, represented by $1385$ solutions. For $252$ showers, two or more solutions are available.
The meteor data available from the MDC are not uniform, varying in the completeness of the averaged shower data. For all showers, the obligatory parameters are: the moment of activity, expressed by the ecliptic longitude of the Sun at the time the shower observation, the geocentric equatorial coordinates of its radiant and the corresponding geocentric velocity. But for many showers the corresponding averaged values of the orbital elements are also given. And for a relatively small number of showers, individual shower member data are available alongside the averaged data.

\begin{figure}[b]
\centering
\includegraphics[width=0.35\textwidth, angle=0]{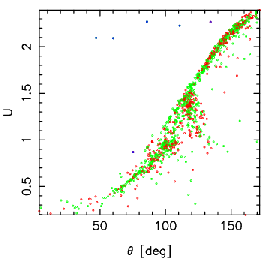}
\caption{Diagram of \"{O}pik's variables $U$ (in the units of the Earth’s velocity) and $\theta$ for meteor data from the MDC database. Red colour indicates points for which ecliptic latitude $|\beta|<10$ [deg]. Blue colour indicates points for which the geocentric data are not consistent with the heliocentric data. These points lie in the area corresponding to open orbits.
}
\label{fig:uteta}
\end{figure}

Hence, in order to make a full comparison of the results obtained in this study, we were limited to a subset of the MDC data allowing the application of each method used. 
We utilized only the data from the 'List of Established Showers' and the 'Working List'. $195$ solutions on these lists were rejected either due to incomplete orbital data or because the orbital eccentricities corresponded to open orbits. We assume that interstellar meteoroid stream solutions are not real and are the result of measurement uncertainties \citep{Hajdukova_Kornos2020}. 
Additional $8$ solutions were removed due to incorrect values of some parameters, e.g. orbital inclinations exceeding $180$ degrees or because of a clear inconsistency between geocentric and heliocentric parameters. This inconsistency is due to the fact that, although orbits with eccentricities $e\geq1$ were removed from our MDC sample, for some meteoroids the values of the \"{O}pik $U$ and $\theta$  variables \citep[see][]{1976iecr.book.....O,1999MNRAS.304..743V}, calculated on the basis of the geocentric parameters clearly fall in the region corresponding to open orbits, see Figure~\ref{fig:uteta}. The \"{O}pik variables, $U$ (in the units of the Earth's velocity) and $\theta$ are defined in an instantaneous geocentric reference frame, and correspond to the geocentric velocity of the meteoroid when encountering the Earth, and to the angular elongation of the meteoroid geocentric radiant from the Earth' apex, respectively. The values of the $\theta$ and $U$ variables shown in Figure \ref{fig:uteta} were calculated using the geocentric coordinates of the radiant and the speed of the meteoroids using the formulas given in \citet{1999MNRAS.304..751J}. 

In the selected MDC dataset, $835$ showers are represented by $1182$ meteor-shower solutions, since the parameters of many showers were determined by more than one author team. $185$ showers are represented by more than one solution. There are $532$ solutions (45.0\% of the total sample examined) which belong to these $185$ MSS. However, it has to be taken into account that we have no assurance that the grouping of the solutions in the MDC is, in all the listed showers, correct, since no objective method was available to detect the duplicates. Table~\ref{tab:MDC-grupy} provides more details on the MSS that occur in the surveyed sample of $1182$ meteor shower solutions.

%
\begin{table}[!t]
\caption{Single- and multi-solution showers in 
the used subset of $1182$ meteor shower data.
The values given in the table are for a sample 
of data taken from the MDC shower database in December 2022.}
\footnotesize
\begin{center}
\begin{tabular}{l r r }
\hline\hline
Showers with one solution only        & 650 \\
Showers with more than one solution   & 185 \\
Showers with two solution             & 106 \\
Showers with tree solutions           & 44 \\
Showers with four solutions           & 13 \\
Showers with five solutions           & 12 \\
Showers with six solutions            & 4 \\
Showers with seven solutions          & 2 \\
Showers with eight solutions          & 1 \\
Showers with nine solutions           & 1 \\
Showers with ten solutions            & 1 \\
Showers with eleven solutions         & 1 \\
Showers with more than eleven solutions  &  0 \\
\hline
\end{tabular}
\end{center}
\label{tab:MDC-grupy}
\normalsize
\end{table}

\begin{figure}
    \centering
    \includegraphics[width=0.30\textwidth, angle=0]{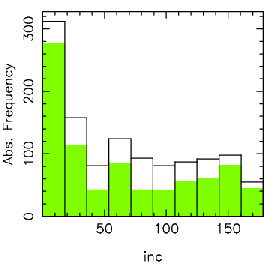}
\caption{Histogram of the inclination 'inc' of $1182$ orbits of the meteor showers used in our study. The green color depicts the orbits for which $q > 0.9\,$au. Orbits with small inclination and high perihelion distances dominate the sample of $1182$ meteor showers listed in the MDC on Dec. 16, 2022.}
\label{fig:inc-q}
\end{figure}

Similarly to the individual meteoroid orbits, the density of meteor showers in near-ecliptical orbits is clearly higher, so we decided (following the idea given by \citet{2001MNRAS.327..623G})  to split the sample of $1182$ shower solutions into two partitions. As can be seen in Figure \ref{fig:inc-q}, the area containing orbits with small inclinations significantly dominates the study sample. It also contains many orbits with perihelion distances greater than $0.9$ [au]. Partition one (P1) contained $498$ shower solutions with orbits with inclinations in the range $0$$-$$40$ [deg], partition P2 $684$ other shower solutions. In the remainder of this paper, we will refer to these subsets of showers as P1- and P2- components, respectively. 

Table~\ref{tab:MDC-all-1} and \ref{tab:MDC-all-2} gives a full list of P1- and P2-  MSS ($86$ and $99$, respectively) that were provided in the MDC database. The minimum number of solutions (duplicates) is, of course, 2, with a maximum of 11 solutions for the Southern Taurids shower (02/STA).~\footnote{
All showers we used were named using the old rules of naming and coding meteoroid streams valid until August 2022, see \citet{2023NewAR..9601671J}.
}
SSS showers are not listed in the Tables~~\ref{tab:MDC-all-1} and \ref{tab:MDC-all-2}.

In the following sections, we describe methodology and the results of the study, which aimed to find new MSS and to assess whether the MSS listed in Table~\ref{tab:MDC-all-1}, Table~\ref{tab:MDC-all-2}  contains duplicates correctly classified.

In Tables~\ref{tab:MDC-all-1},~\ref{tab:MDC-all-2}, we introduced $DH_{Min}$ and $<$$DH$$>$ columns.  The $DH_{Min}$ column contains the smallest threshold values of orbital similarity corresponding to the DH-function \citep{1993Icar..106..603J}, with which all members (all duplicates) of the MSS given in Tables~\ref{tab:MDC-all-1},~\ref{tab:MDC-all-2}, can be identified by cluster analysis using the single linking method. The $<$$DH$$>$ column gives the arithmetic mean of the DH orbital similarity values calculated for all pairs of the given MSS. These values are related to the compactness of the MSS in question. 

From a cursory analysis of the contents of the $DH_{Min}$ column, it is easy to see that for a many of MSS consisted of only two duplicates, we are most likely dealing with cases of false duplicates. Using formulas (21) and (22) given in Table~9 in the \citet{2017P&SS..143...43J}, 
the plausible threshold values of orbital similarity for pairs of orbits taken from a set of ~600 orbits are $0.022$-$0.025$. In contrast, in Table~\ref{tab:MDC-all-1} and \ref{tab:MDC-all-2} we see that for many MSS with only two duplicates the $DH_{Min}$ values are much larger. This observation justifies the revision we have undertaken of the MDC meteor base in terms of the MSS contained therein. Verification and search for duplicates is an important topic, abandoning such research leaves us with artifacts in the MDC database.

\begin{table*}[t]
\caption{ A list of $86$ showers from partition P1- represented in the MDC by multiple solutions. The individual columns give the MDC codes designation of the shower; $N$ --- number of solutions found in the MDC for this group, following the convention adopted in the MDC, twin showers are listed as separate streams, e.g. 008/ORI and 031/ETA;
 $DH_{Min}$ --- the smallest orbital similarity threshold required to identify all members of the MSS's provided in the MDC; $<$$DH$$>$ --- the arithmetic mean value of the DH-function among all members of the MSS.}
\scriptsize
\begin{center}
\begin{tabular}{r l c c c |  r l c c c } 
\hline
\multicolumn{1}{c}{No}&\multicolumn{1}{c}{Shower Code}& \multicolumn{1}{c}{$N$} &
\multicolumn{1}{c}{$DH_{Min}$}& \multicolumn{1}{c}{$<DH>$|} &
\multicolumn{1}{c}{No}&\multicolumn{1}{c}{ Shower Code}& \multicolumn{1}{c}{$N$} &  
\multicolumn{1}{c}{$DH_{Min}$}& \multicolumn{1}{c}{$<$$DH$$>$} \\
\hline
\hline
  1 &  0001/00/CAP &    9 &  0.062 &  0.048 &  51 &  0197/00/AUD &    2 &  0.316 &  0.316 \\
  2 &  0002/00/STA &   11 &  0.182 &  0.205 &  52 &  0199/00/ADC &    2 &  0.107 &  0.107 \\
  3 &  0003/00/SIA &    2 &  0.065 &  0.065 &  53 &  0202/00/ZCA &    2 &  0.434 &  0.434 \\
  4 &  0004/00/GEM &    5 &  0.020 &  0.022 &  54 &  0212/01/KLE &    3 &  0.303 &  0.221 \\
  5 &  0005/00/SDA &    8 &  0.078 &  0.097 &  55 &  0215/00/NPI &    4 &  0.170 &  0.138 \\
  6 &  0009/00/DRA &    4 &  0.186 &  0.126 &  56 &  0216/00/SPI &    5 &  0.460 &  0.255 \\
  7 &  0011/00/EVI &    3 &  0.322 &  0.284 &  57 &  0219/00/SAR &    4 &  0.373 &  0.398 \\
  8 &  0012/00/KCG &    4 &  0.117 &  0.114 &  58 &  0220/00/NDR &    2 &  0.144 &  0.144 \\
  9 &  0017/00/NTA &   10 &  0.240 &  0.157 &  59 &  0221/00/DSX &    5 &  0.115 &  0.082 \\
 10 &  0018/00/AND &    3 &  0.044 &  0.048 &  60 &  0233/00/OCC &    2 &  0.160 &  0.160 \\
 11 &  0019/00/MON &    5 &  0.044 &  0.047 &  61 &  0250/00/NOO &    3 &  0.053 &  0.066 \\
 12 &  0021/00/AVB &    3 &  0.124 &  0.097 &  62 &  0253/00/CMI &    2 &  0.347 &  0.347 \\
 13 &  0025/00/NOA &    2 &  0.210 &  0.210 &  63 &  0254/00/PHO &    2 &  0.199 &  0.199 \\
 14 &  0026/00/NDA &    7 &  0.131 &  0.106 &  64 &  0256/00/ORN &    4 &  0.201 &  0.192 \\
 15 &  0028/00/SOA &    2 &  0.113 &  0.113 &  65 &  0257/00/ORS &    5 &  0.123 &  0.161 \\
 16 &  0033/00/NIA &    5 &  0.158 &  0.144 &  66 &  0338/01/OER &    2 &  0.023 &  0.023 \\
 17 &  0047/00/DLI &    2 &  0.096 &  0.096 &  67 &  0343/02/HVI &    6 &  0.103 &  0.050 \\
 18 &  0061/00/TAH &    2 &  0.071 &  0.071 &  68 &  0388/00/CTA &    3 &  0.127 &  0.133 \\
 19 &  0069/00/SSG &    3 &  0.214 &  0.168 &  69 &  0390/00/THA &    2 &  0.135 &  0.135 \\
 20 &  0076/00/KAQ &    2 &  0.142 &  0.142 &  70 &  0446/00/DPC &    2 &  0.010 &  0.010 \\
 21 &  0088/00/ODR &    2 &  0.362 &  0.362 &  71 &  0451/00/CAM &    3 &  0.107 &  0.086 \\
 22 &  0096/01/NCC &    6 &  0.162 &  0.147 &  72 &  0456/00/MPS &    4 &  0.041 &  0.042 \\
 23 &  0097/00/SCC &    4 &  0.218 &  0.207 &  73 &  0459/00/JEO &    3 &  0.050 &  0.041 \\
 24 &  0100/00/XSA &    2 &  0.209 &  0.209 &  74 &  0460/00/LOP &    3 &  0.044 &  0.052 \\
 25 &  0112/01/NDL &    2 &  0.070 &  0.070 &  75 &  0464/00/KLY &    3 &  0.030 &  0.031 \\
 26 &  0113/00/SDL &    2 &  0.172 &  0.172 &  76 &  0470/00/AMD &    3 &  0.075 &  0.065 \\
 27 &  0115/00/DCS &    5 &  0.257 &  0.264 &  77 &  0486/00/NZP &    2 &  0.041 &  0.041 \\
 28 &  0121/00/NHY &    3 &  0.269 &  0.253 &  78 &  0490/00/DGE &    2 &  0.149 &  0.149 \\
 29 &  0124/00/SVI &    2 &  0.226 &  0.226 &  79 &  0505/00/AIC &    3 &  0.117 &  0.090 \\
 30 &  0127/00/MCA &    2 &  0.041 &  0.041 &  80 &  0525/00/ICY &    2 &  0.031 &  0.031 \\
 31 &  0128/00/MKA &    2 &  0.315 &  0.315 &  81 &  0538/00/FFA &    2 &  0.092 &  0.092 \\
 32 &  0133/00/PUM &    2 &  0.139 &  0.139 &  82 &  0644/00/JLL &    2 &  0.256 &  0.256 \\
 33 &  0144/00/APS &    3 &  0.094 &  0.072 &  83 &  0651/00/OAV &    2 &  0.096 &  0.096 \\
 34 &  0150/01/SOP &    2 &  0.343 &  0.343 &  84 &  0709/00/LCM &    2 &  0.142 &  0.142 \\
 35 &  0152/00/NOC &    4 &  0.644 &  0.447 &  85 &  0757/00/CCY &    4 &  0.035 &  0.033 \\
 36 &  0153/00/OCE &    3 &  0.081 &  0.074 &  86 &  1046/00/PIS &    2 &  0.078 &  0.078 \\
 37 &  0154/00/DEA &    2 &  0.178 &  0.178 &     &              &      &        &        \\
 38 &  0155/00/NMA &    2 &  0.160 &  0.160 &     &              &      &        &        \\
 39 &  0156/00/SMA &    3 &  0.362 &  0.305 &     &              &      &        &        \\
 40 &  0164/00/NZC &    5 &  0.225 &  0.151 &     &              &      &        &        \\
 41 &  0165/01/SZC &    3 &  0.510 &  0.376 &     &              &      &        &        \\
 42 &  0167/00/NSS &    2 &  0.341 &  0.341 &     &              &      &        &        \\
 43 &  0170/00/JBO &    2 &  0.220 &  0.220 &     &              &      &        &        \\
 44 &  0171/00/ARI &    5 &  0.090 &  0.088 &     &              &      &        &        \\
 45 &  0172/00/ZPE &    4 &  0.075 &  0.096 &     &              &      &        &        \\
 46 &  0173/00/BTA &    4 &  0.191 &  0.194 &     &              &      &        &        \\
 47 &  0179/00/SCA &    2 &  0.234 &  0.234 &     &              &      &        &        \\
 48 &  0186/00/EUM &    2 &  0.194 &  0.194 &     &              &      &        &        \\
 49 &  0188/00/XRI &    3 &  0.534 &  0.496 &     &              &      &        &        \\
 50 &  0189/00/DMC &    2 &  0.454 &  0.454 &     &              &      &        &        \\
 \hline
\end{tabular}
\end{center}
\label{tab:MDC-all-1}
\normalsize
\end{table*}
\begin{table*}[t]
\caption{ A list of $99$ showers represented in the MDC, partition P2- by multiple solutions. See  description of columns in Table~\ref{tab:MDC-all-1}.
}
\scriptsize
\begin{center}
\begin{tabular}{r l c c c |  r l c c c } 
\hline
\multicolumn{1}{c}{No}&\multicolumn{1}{c}{Shower Code}& \multicolumn{1}{c}{$N$} &
\multicolumn{1}{c}{$DH_{Min}$}& \multicolumn{1}{c}{$<DH>$|} &
\multicolumn{1}{c}{No}&\multicolumn{1}{c}{ Shower Code}& \multicolumn{1}{c}{$N$} &  
\multicolumn{1}{c}{$DH_{Min}$}& \multicolumn{1}{c}{$<$$DH$$>$} \\
\hline
\hline
   1 &  0006/00/LYR &    6 &  0.045 &  0.041 &   51 &  0404/00/GUM &    3 &  0.132 &  0.097 \\
  2 &  0007/00/PER &    3 &  0.011 &  0.010 &	52 &  0428/00/DSV &    2 &  0.067 &  0.067 \\
  3 &  0008/00/ORI &    5 &  0.062 &  0.049 &	53 &  0429/00/ACB &    3 &  0.041 &  0.048 \\
  4 &  0010/00/QUA &    6 &  0.050 &  0.050 &	54 &  0431/00/JIP &    2 &  0.062 &  0.062 \\
  5 &  0013/00/LEO &    7 &  0.158 &  0.129 &	55 &  0439/00/ASX &    2 &  0.077 &  0.077 \\
  6 &  0015/03/URS &    3 &  0.149 &  0.118 &	56 &  0444/00/ZCS &    2 &  0.045 &  0.045 \\
  7 &  0016/00/HYD &    3 &  0.095 &  0.106 &	57 &  0450/00/AED &    2 &  0.061 &  0.061 \\
  8 &  0020/04/COM &    3 &  0.259 &  0.207 &	58 &  0458/00/JEC &    3 &  0.081 &  0.086 \\
  9 &  0022/00/LMI &    2 &  0.037 &  0.037 &	59 &  0465/00/AXC &    3 &  0.064 &  0.083 \\
 10 &  0023/00/EGE &    2 &  0.099 &  0.099 &	60 &  0466/01/AOC &    2 &  0.086 &  0.086 \\
 11 &  0031/00/ETA &    5 &  0.109 &  0.104 &	61 &  0479/00/SOO &    3 &  0.072 &  0.065 \\
 12 &  0032/00/DLM &    2 &  0.321 &  0.321 &	62 &  0480/00/TCA &    3 &  0.066 &  0.056 \\
 13 &  0040/00/ZCY &    2 &  0.259 &  0.259 &	63 &  0481/00/OML &    3 &  0.031 &  0.026 \\
 14 &  0093/00/VEL &    2 &  0.412 &  0.412 &	64 &  0488/00/NSU &    2 &  0.041 &  0.041 \\
 15 &  0105/00/OCN &    2 &  0.134 &  0.134 &	65 &  0491/00/DCC &    2 &  0.093 &  0.093 \\
 16 &  0106/00/API &    2 &  0.329 &  0.329 &	66 &  0494/00/DEL &    2 &  0.028 &  0.028 \\
 17 &  0107/00/DCH &    3 &  0.358 &  0.388 &	67 &  0497/01/DAB &    2 &  0.048 &  0.048 \\
 18 &  0108/00/BTU &    2 &  0.344 &  0.344 &	68 &  0498/00/DMH &    2 &  0.096 &  0.096 \\
 19 &  0110/00/AAN &    5 &  0.109 &  0.094 &	69 &  0500/00/JPV &    3 &  0.114 &  0.105 \\
 20 &  0118/00/GNO &    2 &  0.798 &  0.798 &	70 &  0502/00/DRV &    3 &  0.042 &  0.039 \\
 21 &  0151/00/EAU &    2 &  0.155 &  0.155 &	71 &  0506/00/FEV &    2 &  0.055 &  0.055 \\
 22 &  0175/02/JPE &    4 &  0.103 &  0.094 &	72 &  0507/00/UAN &    2 &  0.326 &  0.326 \\
 23 &  0183/00/PAU &    3 &  0.294 &  0.312 &	73 &  0510/00/JRC &    2 &  0.068 &  0.068 \\
 24 &  0184/02/GDR &    2 &  0.006 &  0.006 &	74 &  0512/00/RPU &    2 &  0.370 &  0.370 \\
 25 &  0187/00/PCA &    3 &  0.314 &  0.280 &	75 &  0519/00/BAQ &    2 &  0.041 &  0.041 \\
 26 &  0191/00/ERI &    2 &  0.049 &  0.049 &	76 &  0520/00/MBC &    2 &  0.041 &  0.041 \\
 27 &  0319/00/JLE &    3 &  0.143 &  0.122 &	77 &  0523/00/AGC &    2 &  0.030 &  0.029 \\
 28 &  0320/00/OSE &    2 &  0.281 &  0.280 &	78 &  0526/00/SLD &    3 &  0.032 &  0.033 \\
 29 &  0321/00/TCB &    2 &  0.089 &  0.089 &	79 &  0529/00/EHY &    3 &  0.041 &  0.046 \\
 30 &  0322/00/LBO &    2 &  0.072 &  0.072 &	80 &  0530/01/ECV &    2 &  0.069 &  0.069 \\
 31 &  0323/00/XCB &    4 &  0.100 &  0.120 &	81 &  0531/00/GAQ &    3 &  0.253 &  0.238 \\
 32 &  0324/00/EPR &    2 &  0.250 &  0.250 &	82 &  0533/00/JXA &    3 &  0.037 &  0.040 \\
 33 &  0326/00/EPG &    2 &  0.175 &  0.175 &	83 &  0537/00/KAU &    2 &  0.141 &  0.141 \\
 34 &  0327/00/BEQ &    2 &  0.377 &  0.377 &	84 &  0545/00/XCA &    2 &  0.135 &  0.135 \\
 35 &  0330/00/SSE &    3 &  0.157 &  0.121 &	85 &  0549/00/FAN &    2 &  0.105 &  0.105 \\
 36 &  0331/00/AHY &    3 &  0.050 &  0.045 &	86 &  0552/00/PSO &    2 &  0.250 &  0.250 \\
 37 &  0333/00/OCU &    2 &  0.100 &  0.100 &	87 &  0555/00/OCP &    2 &  0.282 &  0.282 \\
 38 &  0334/01/DAD &    2 &  0.110 &  0.110 &	88 &  0558/00/TSM &    2 &  0.076 &  0.076 \\
 39 &  0337/02/NUE &    2 &  0.221 &  0.221 &	89 &  0563/00/DOU &    2 &  0.030 &  0.030 \\
 40 &  0340/01/TPY &    2 &  0.008 &  0.008 &	90 &  0567/00/XHY &    2 &  0.069 &  0.069 \\
 41 &  0341/02/XUM &    2 &  0.025 &  0.025 &	91 &  0569/00/OHY &    2 &  0.082 &  0.082 \\
 42 &  0346/01/XHE &    3 &  0.057 &  0.052 &	92 &  0570/00/FBH &    2 &  0.024 &  0.024 \\
 43 &  0347/00/BPG &    2 &  0.230 &  0.230 &	93 &  0574/00/GMA &    2 &  0.480 &  0.480 \\
 44 &  0348/00/ARC &    2 &  0.096 &  0.096 &	94 &  0746/00/EVE &    2 &  0.085 &  0.085 \\
 45 &  0362/00/JMC &    2 &  0.125 &  0.125 &	95 &  0842/00/CRN &    2 &  0.130 &  0.130 \\
 46 &  0372/00/PPS &    2 &  0.350 &  0.350 &	96 &  1044/00/EPU &    2 &  0.034 &  0.034 \\
 47 &  0386/00/OBC &    2 &  0.266 &  0.266 &	97 &  1048/00/JAS &    2 &  0.150 &  0.150 \\
 48 &  0392/00/NID &    2 &  0.317 &  0.317 &	98 &  1049/00/DIU &    3 &  0.147 &  0.143 \\
 49 &  0394/00/ACA &    2 &  0.070 &  0.070 &	99 &  1106/00/GAD &    2 &  0.069 &  0.069 \\
 50 &  0398/00/DCM &    2 &  0.203 &  0.203 &       &              &      &        &         \\
\hline
\end{tabular}
\end{center}
\label{tab:MDC-all-2}
\normalsize
\end{table*}
%
\section{Methodology for finding duplicates in the MDC}
\label{methodology}
A disadvantage of the approaches applied in this study is that we most often have to use individually averaged values of shower parameters. It means that the multidimensional averaging problem was substituted by a series of one dimensional problems. It is well known \citep[see][]{1986PhDT........61J, 1999md98.conf..199V, 1999SoSyR..33..302V, 2001ESASP.495...33W, 2006MNRAS.371.1367J, 2010pim8.conf...91J} that individual averaging can introduce some level of inconsistency between the geocentric and heliocentric parameters, and even between the heliocentric parameters themselves. For example the individually averaged orbital elements $q$, $a$, $e$ may not necessarily satisfy the known formula $q = a(1 - e)$. The same problem occurs when the medians of the meteoroid stream parameters are used. 

In our reduced sample of $1182$ meteor shower data, for each shower averaged geocentric and corresponding heliocentric parameters are given. Assuming that the impact of individual averaging of meteoroid parameters is not a significant obstacle to our idea, using these parameters, a search for duplicates among the MDC shower data can be done in a manner analogous to the search for streams among orbits of individual meteoroids. 
For this purpose, it is sufficient to use the cluster analysis methods, which have been used for years in a search for meteoroid streams, with the difference that, this time, the identification of groups in the MDC will be made among the mean radiants or orbital elements or their combinations. As a result, the groups (MSS) thus identified will consist of duplicates (the shower multi-solutions) we are looking for.

We performed several cluster analyses among $494$ solutions in the P1 partition, and among $688$  solutions collected in the P2 partition.  We decided to search the two partitions separately so that the cluster analysis would be performed with threshold values of orbital similarity determined for each partition separately. This solution reduced the unfavourable influence coming from the domination of orbits with relatively small inclinations to the ecliptic in the studied sample of $1182$ showers. 

We used a single linkage method (a variant of the general hierarchical cluster analysis method) successfully used  by a number of authors for the meteoroid stream identification or searching for grouping among the asteroids: \citet{1963SCoA....7..261S, 1971SCoA...12...14L, 1990AJ....100.2030Z,  1992acm..proc..363L, 1993mtpb.conf..269J, 1994AJ....107..772Z, 1995Icar..116..291Z, 1999md98.conf..307J, 2003MNRAS.344..665J, 2010MNRAS.404..867J, 2020MNRAS.494..680J}. 
\subsection{Method-I --- cluster analysis among the orbital data}
\label{Method-1}
Our first  approach, (method-I), takes advantage of the orbital similarity of meteoroids, calculated by the D-function; in the cluster analysis, the single linking method was used, \citep[see][]{1963SCoA....7..261S,1997A&A...320..631J}.
The orbital D-distances were calculated by the orbital similarity function $DH$ described in \citet[][]{1993Icar..106..603J, 2019msme.book..210W}. The orbital similarity thresholds were found separately for groups of $M$$=$$2,3,4,5,...$ members; threshold for $M=10$ was applied for all groups for which $M>10$. 
All thresholds corresponded to a low probability (less than 1 per cent) of chance grouping. Threshold values were calculated using the method presented in \citet{2020MNRAS.494..680J}; for both P1- and P2- partitions  they are listed in Table \ref{tab:thresholds}. 

In the cluster analysis, we restricted the individual thresholds to the maximum value corresponding to $M=10$. Such a decision is justified by the fact that only one of the MSS in Table~\ref{tab:MDC-grupy} contain more than $10$ duplicates, and our preliminary calculations showed that the threshold values of orbital similarity for the $DH$ function determined for $M>10$ were too large and led to results that were difficult to interpret. This may have to do with the limitations of the statistical approach for estimating threshold values as mentioned in the \citet{2003MNRAS.344..665J} paper. 
As a reminder, the orbital similarity thresholds used in method-I are only applicable in the cluster analysis performed by the single linking method. Of course, this does not apply to groups with only two members. The threshold values given in Table ~\ref{tab:thresholds} corresponding to $M=2$ do not depend on the choice of cluster analysis algorithm.
\begin{table}
\caption{The orbital $DH$ similarity thresholds and their uncertainties applied in the search for duplicates among MDC shower data. On the left, threshold values for $498$ orbits (P1- partition), on the right for $684$ orbits, (P2-partition). The thresholds correspond to the reliability level $>99\%$ for  each group of $M$ members and the distance functions $DH$.  The values provided in this table are closely related with the single-linkage method and the size of the MDC sample used in this study.}
\begin{center}
\scriptsize
\begin{tabular}{ccc|cc}
\hline
\multicolumn{1}{c}{ } &  \multicolumn{2}{c}{Partition 1} &
                         \multicolumn{2}{|c}{Partition 2}\\
\hline
\multicolumn{1}{c|}{M} &  \multicolumn{2}{c}{$D{H} \pm \sigma$} &
                         \multicolumn{2}{|c}{$D{H} \pm \sigma$}\\
\hline
\hline	
   2 &   0.039  &  0.001  &   0.059  &  0.001 \\
   3 &   0.079  &  0.001  &   0.136  &  0.002 \\
   4 &   0.099  &  0.001  &   0.182  &  0.002 \\
   5 &   0.112  &  0.001  &   0.208  &  0.002 \\
   6 &   0.121  &  0.001  &   0.226  &  0.002 \\
   7 &   0.128  &  0.001  &   0.244  &  0.002 \\
   8 &   0.134  &  0.001  &   0.257  &  0.002 \\
   9 &   0.139  &  0.001  &   0.267  &  0.002 \\
  10 &   0.142  &  0.001  &   0.274  &  0.002 \\
\hline
\hline
\end{tabular}
\end{center}
\label{tab:thresholds}
\normalsize
\end{table}

We note that method to identify the duplicates will probably have to include a cluster analysis e.g.  linking procedure (single or other linking), since a linking of more than two solutions may occur non-trivial. 
Let us consider, for example, three solutions, A, B, and C. With the help of a method, we find that B is the duplicate of A, therefore the solution B should be one of the solutions of shower solution A is belonging to. Further, we find that C is the duplicate of B, therefore solution C should be the solution of the same shower as B, i.e. the shower containing A. But, applying the method to pair A and C, we find that C is not a duplicate of A, therefore a controversy, that C should not be the solution of shower containing A, occurs. The single linking method can solve this problem.
\subsection{Method-II --- maximum-sigma approach}
\label{Method-2}
In the second method  (method-II), we used the idea of similarity between two showers proposed by \citet{2020eMetN...5...93K}. Koseki compared three shower parameters: the ecliptic coordinates of the shower radiant (the Sun-centered longitude and the radiant latitude)  and the ecliptic longitude of the Sun at the time of shower activity. To assess the similarity of the two showers, Koseki calculated the differences between the relevant parameters and compared them with selected critical values. However, the author did not explain why he chose the critical values and not others, as well as he did not apply a cluster analysis to the whole data set, limiting his research to comparing each time only two showers. Which in our view is a very limited approach. 
Therefore,  in this study, we decided to extend Koseki's approach by comparing both geocentric and heliocentric shower parameters, as well as performing cluster analysis. Moreover, in our approach, we have justified the choice of such and not other critical values of differences of the compared quantities. 

When we want to determine whether a shower solution is an autonomous or a duplicate, we should be able to justify the difference between the parameters of that shower and the others in the database.
So, we examine a set of mean parameters and in case of an autonomous shower, we demand that the compared solutions must significantly differ at least in one of these parameters.

To evaluate whether the difference among two values of the meteor shower parameter is significant, we propose similar approach which was recently applied in cosmology to reason that the Hubble constant determined by two methods is actually different, \citep[see e.g.][]{Jones_etal2020, DiValentino_etal2021, Perivolaropous_Skara2021}. Namely, the so-called three-sigma rule of thumb (or 3-$\sigma$ rule). In empirical science, this rule expresses a conventional heuristic that nearly all values lie within three standard deviations of the mean \citep{statistics2003}. With respect to meteor showers, it can therefore be argued that if the critical difference of two determinations of some shower's parameter lies outside the $\pm 3$-$\sigma$ interval, then the two meteor shower solutions being compared are not duplicates. 

Unfortunately, the method based on the 3-$\sigma$ difference of a parameter can be used only for a small fraction of the meteor showers, since the determination errors of their parameters are unknown for majority solutions in the MDC database. This circumstance forces us to propose and use a less exact, but generally applicable method, which we refer, hereafter, as the ``maximum-$\sigma$ method'' and which is outlined below.

 In December 2022, in addition to the shower average parameters, the MDC contained so-called LookUp Tables \citep[see][]{2023A&A...671A.155H} available for $127$ showers. The contents of the LookUp Tables made it possible to calculate the standard deviations of the shower parameters of interest and, in a further step, to construct the cumulative distribution of these 1-$\sigma$ values.
\begin{figure*}[!t]
  \centerline{\includegraphics[width=0.21\textwidth, angle=-90]{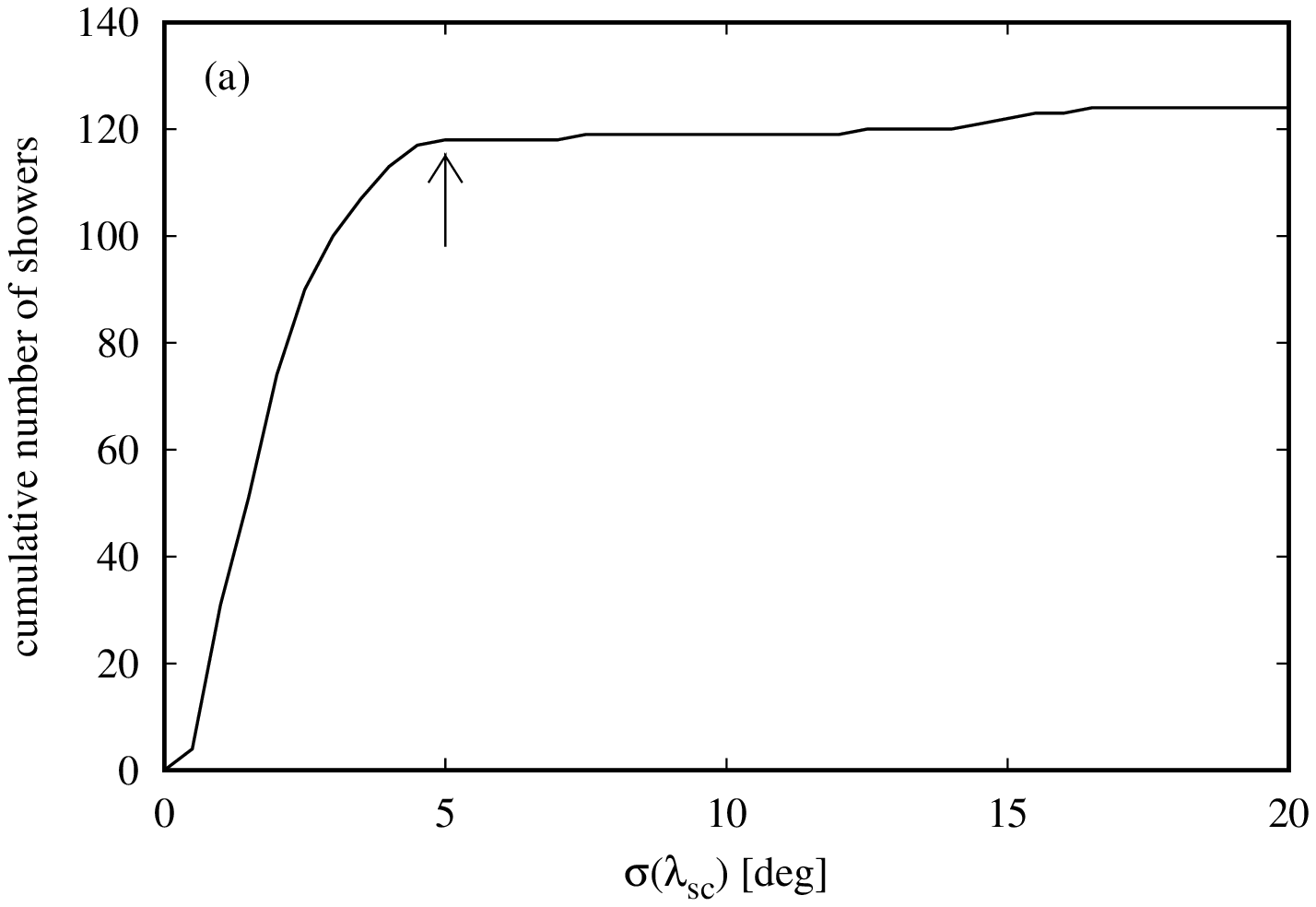}
              \includegraphics[width=0.21\textwidth, angle=-90]{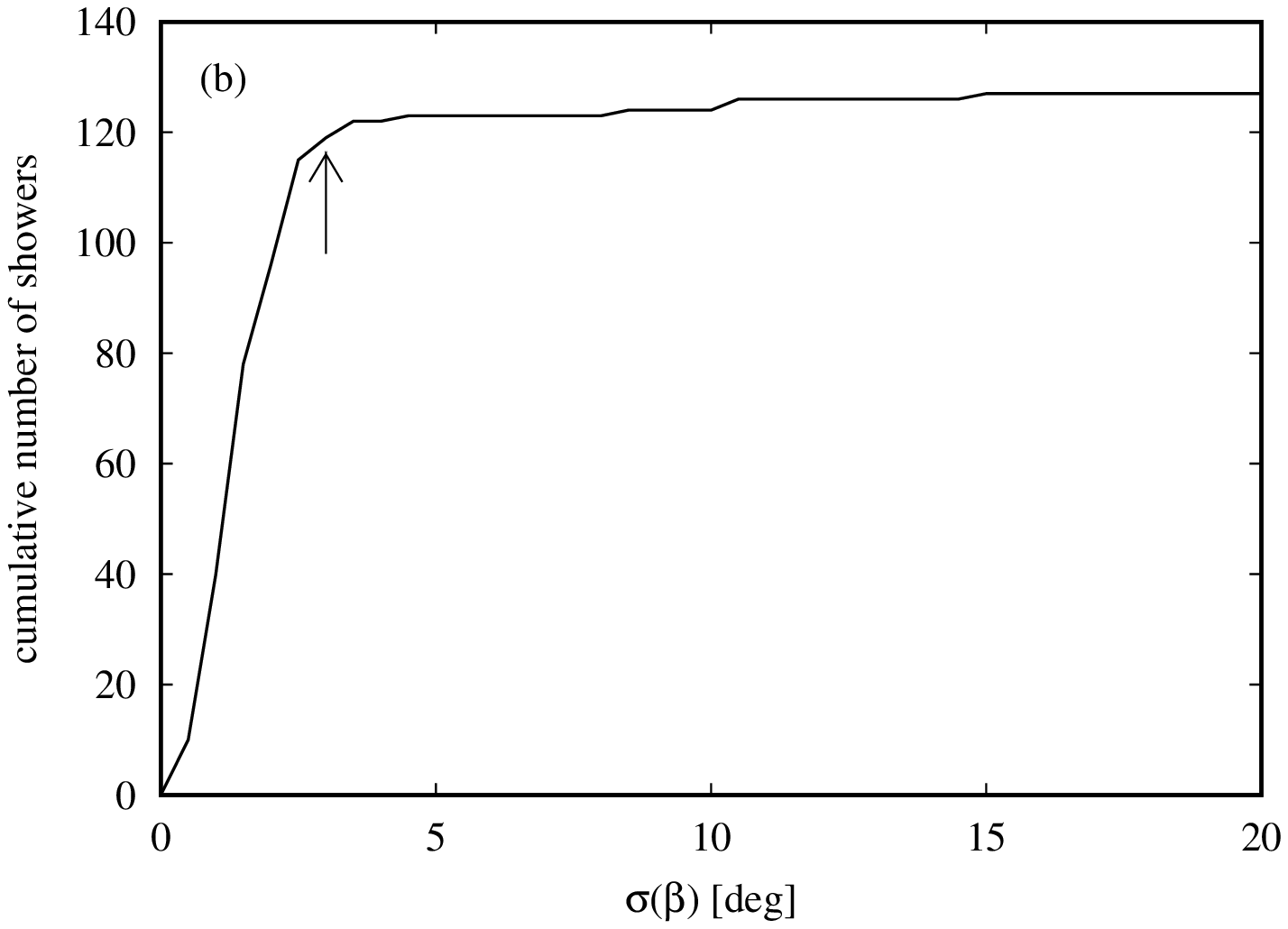}
              \includegraphics[width=0.21\textwidth, angle=-90]{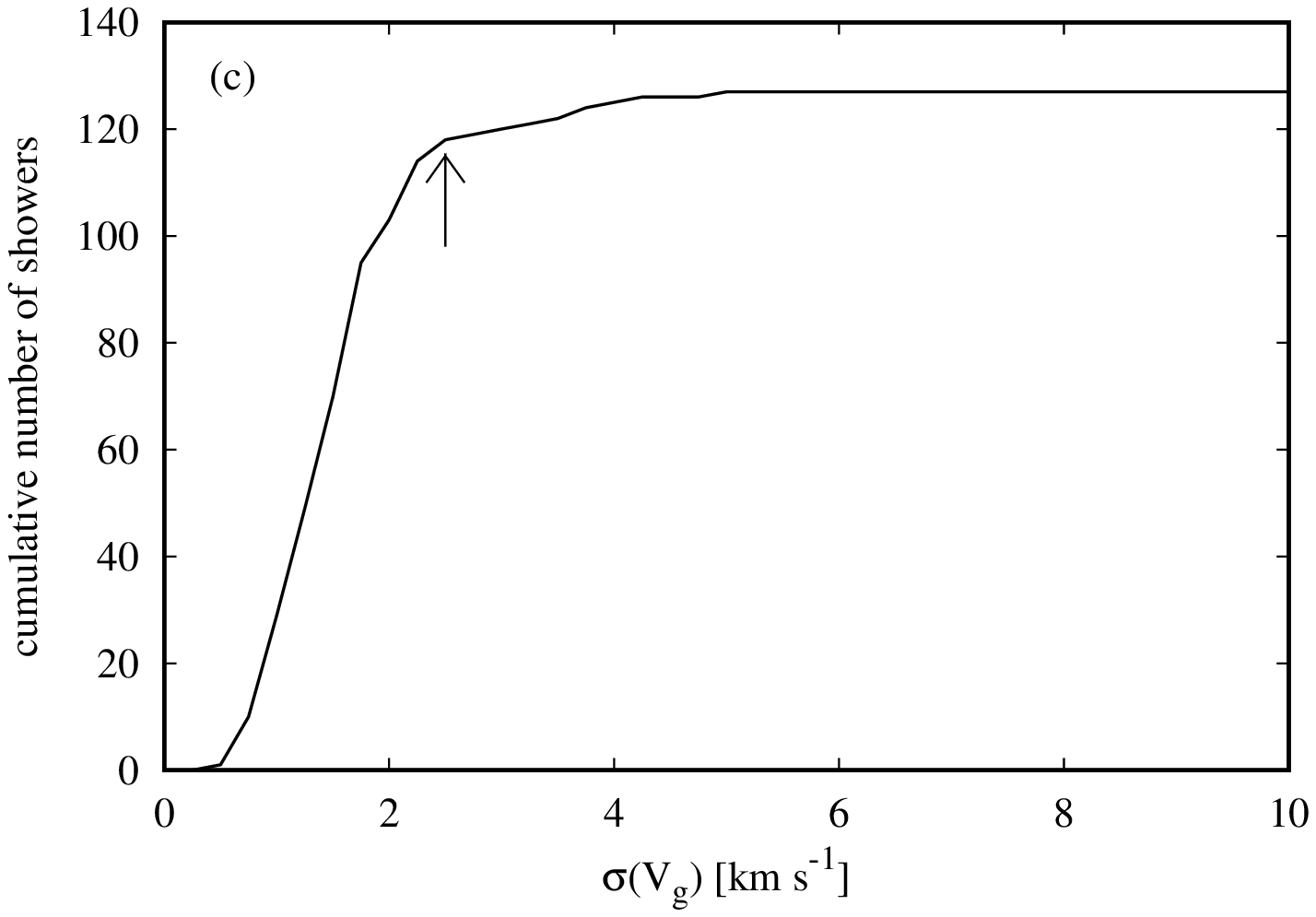}}      
  \centerline{\includegraphics[width=0.21\textwidth, angle=-90]{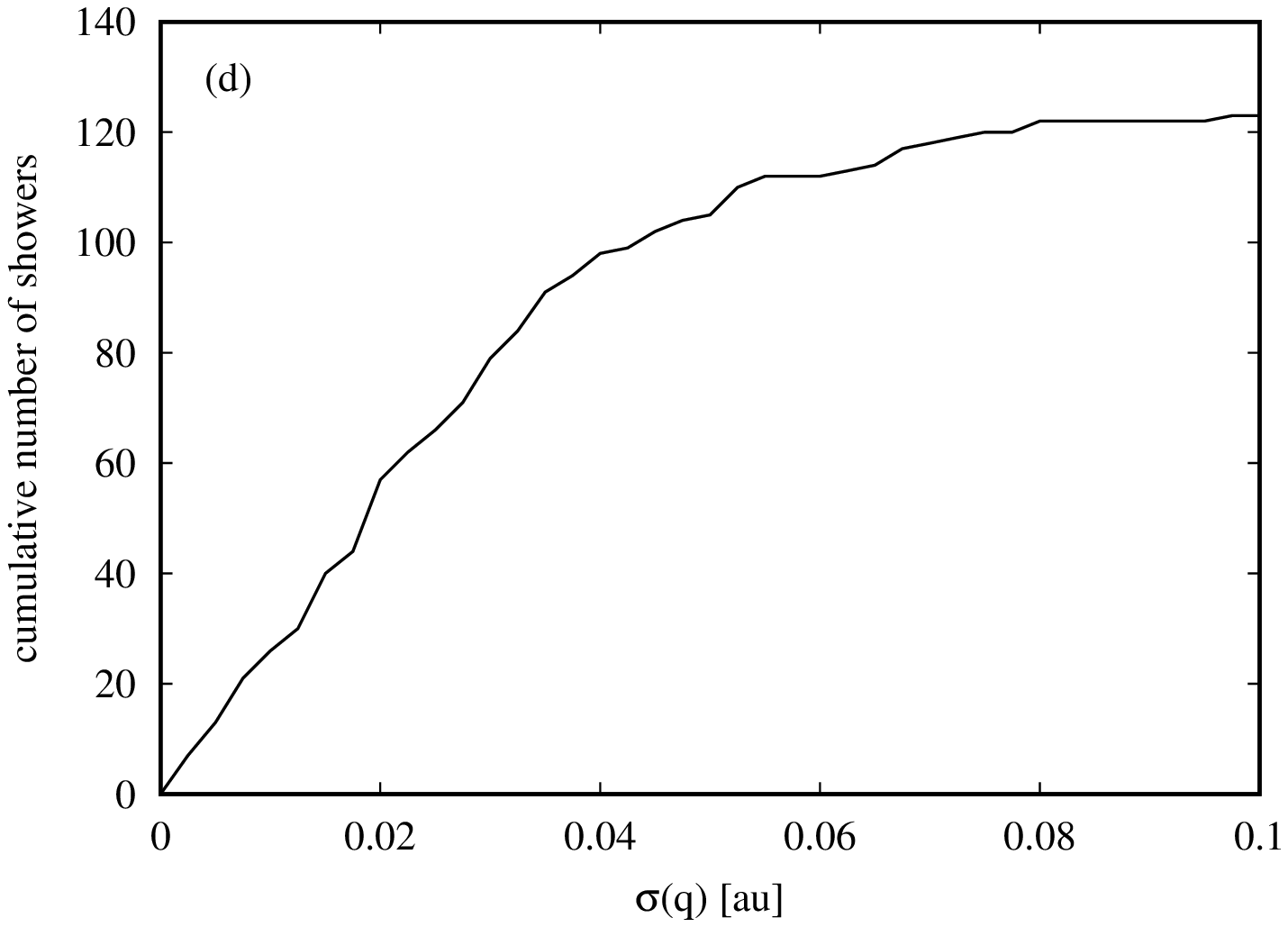}
              \includegraphics[width=0.21\textwidth, angle=-90]{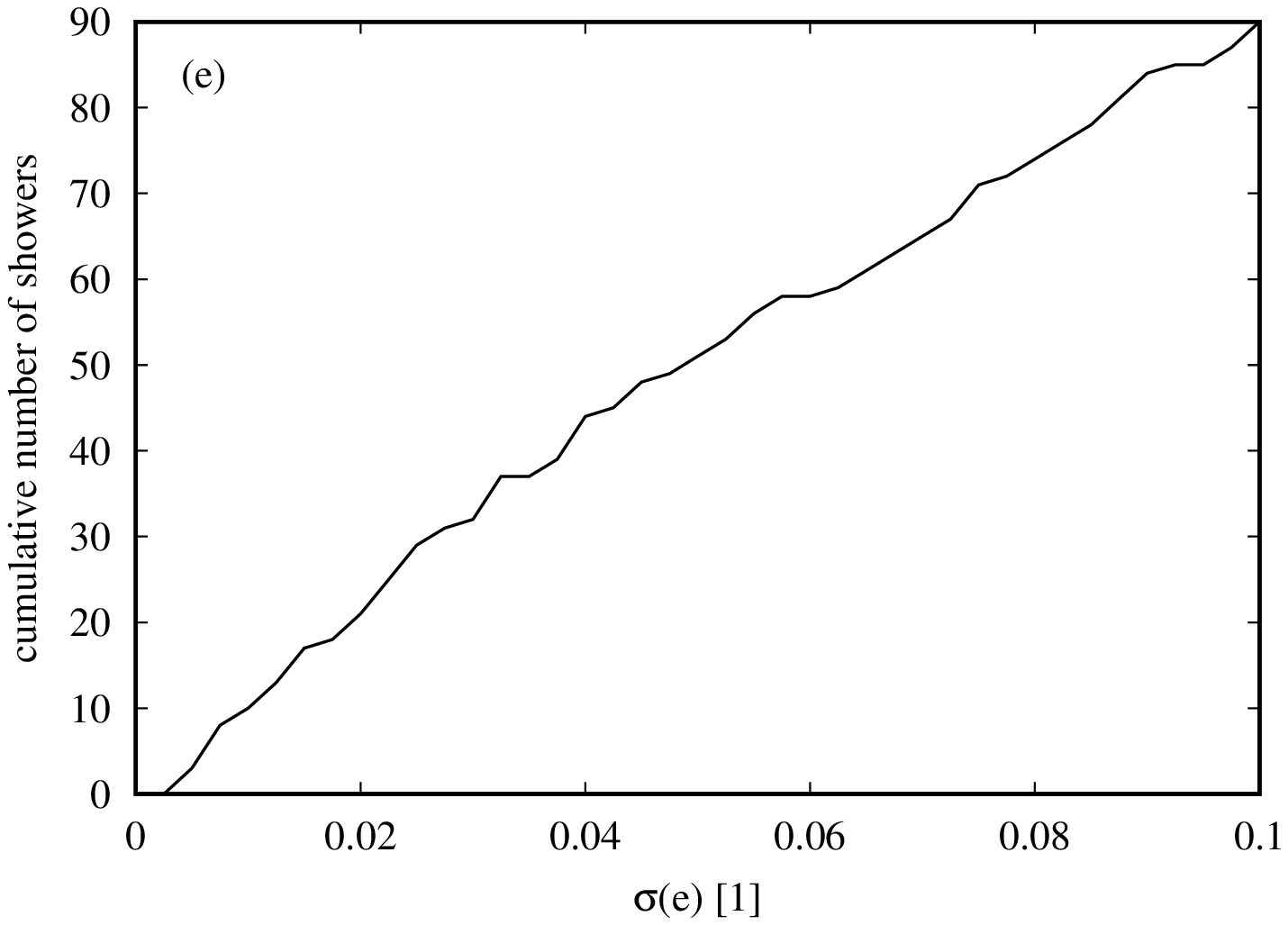}
              \includegraphics[width=0.21\textwidth, angle=-90]{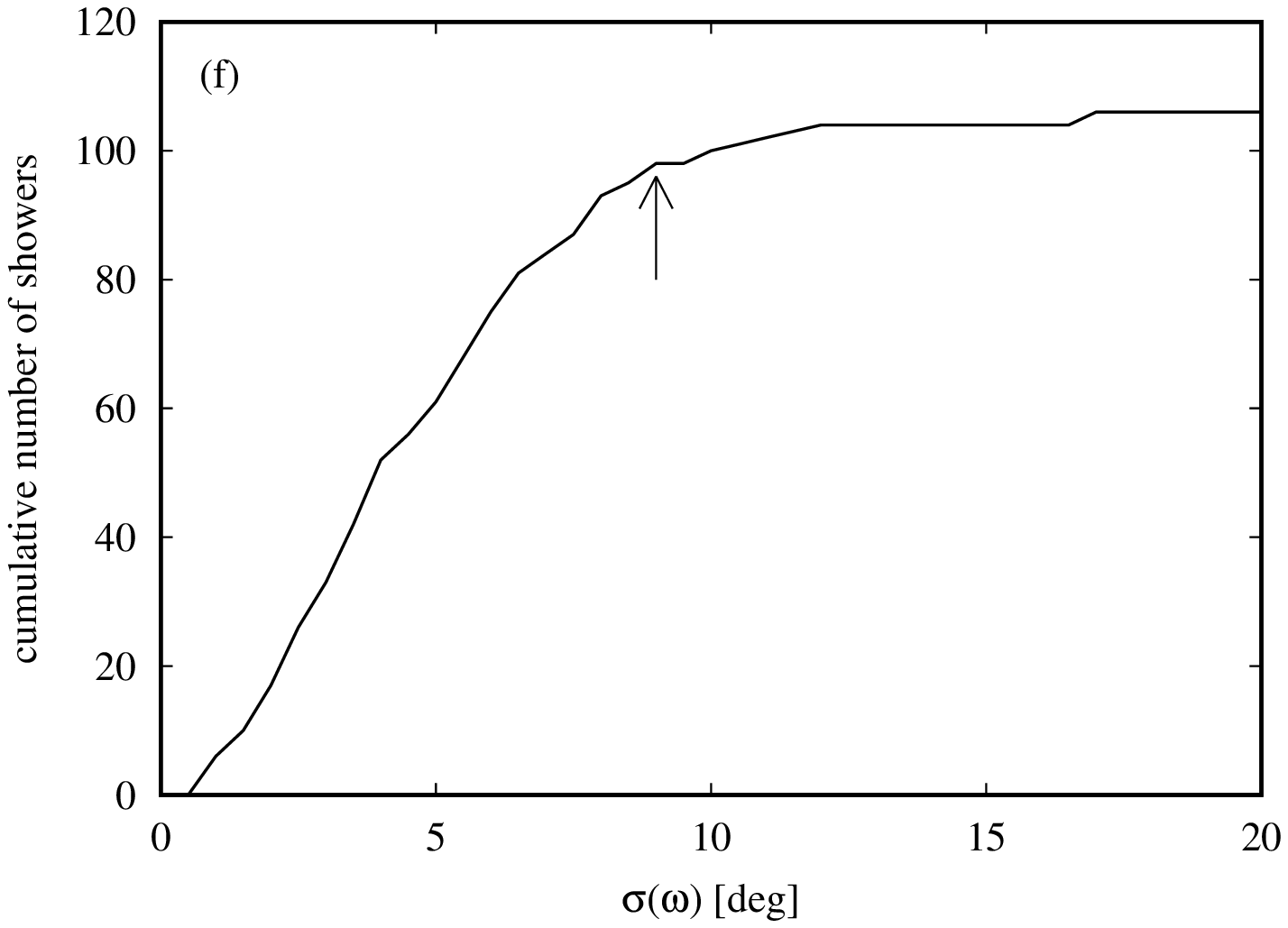}}
   \centerline{\includegraphics[width=0.21\textwidth, angle=-90]{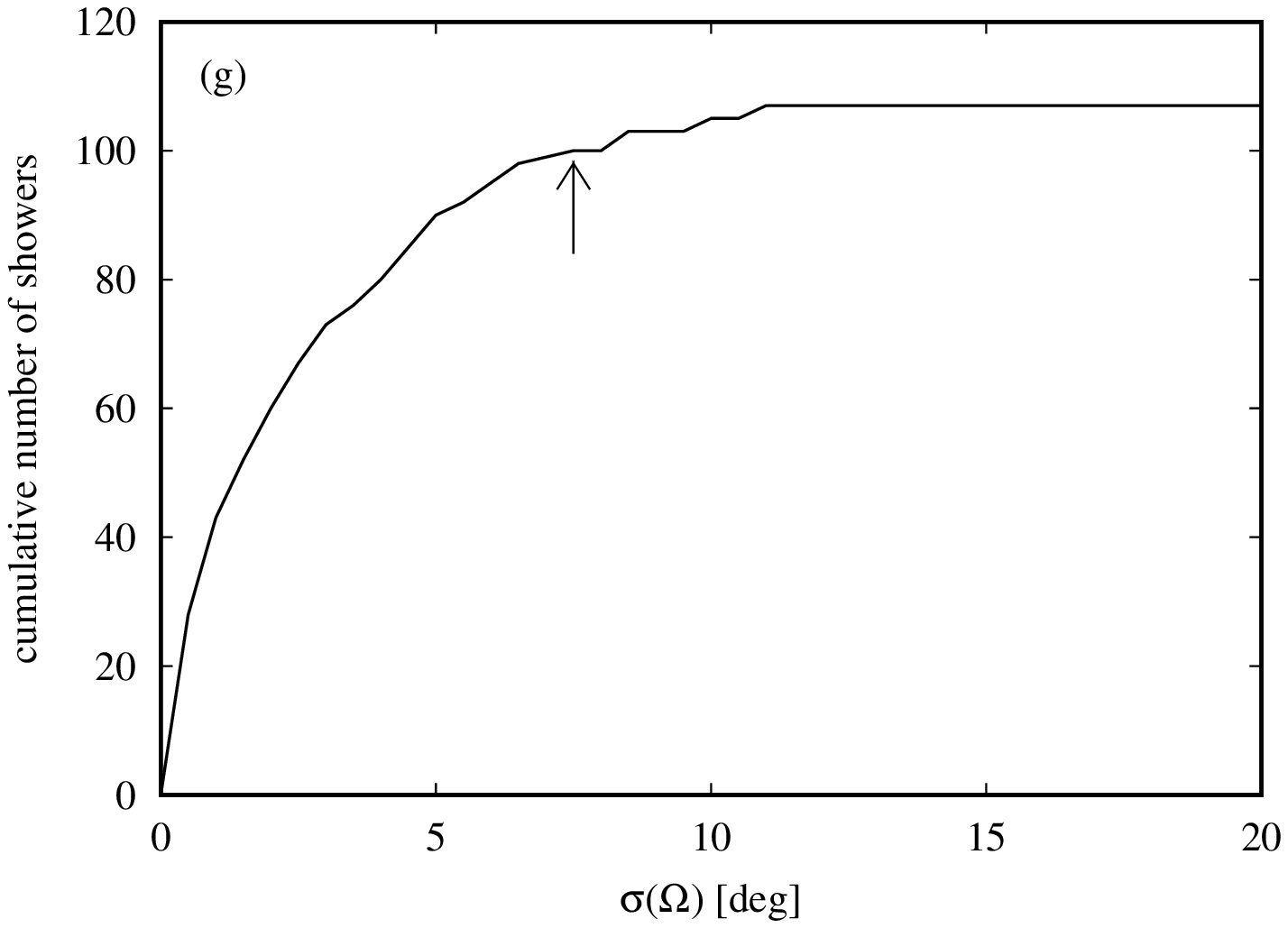}                          
               \includegraphics[width=0.21\textwidth, angle=-90]{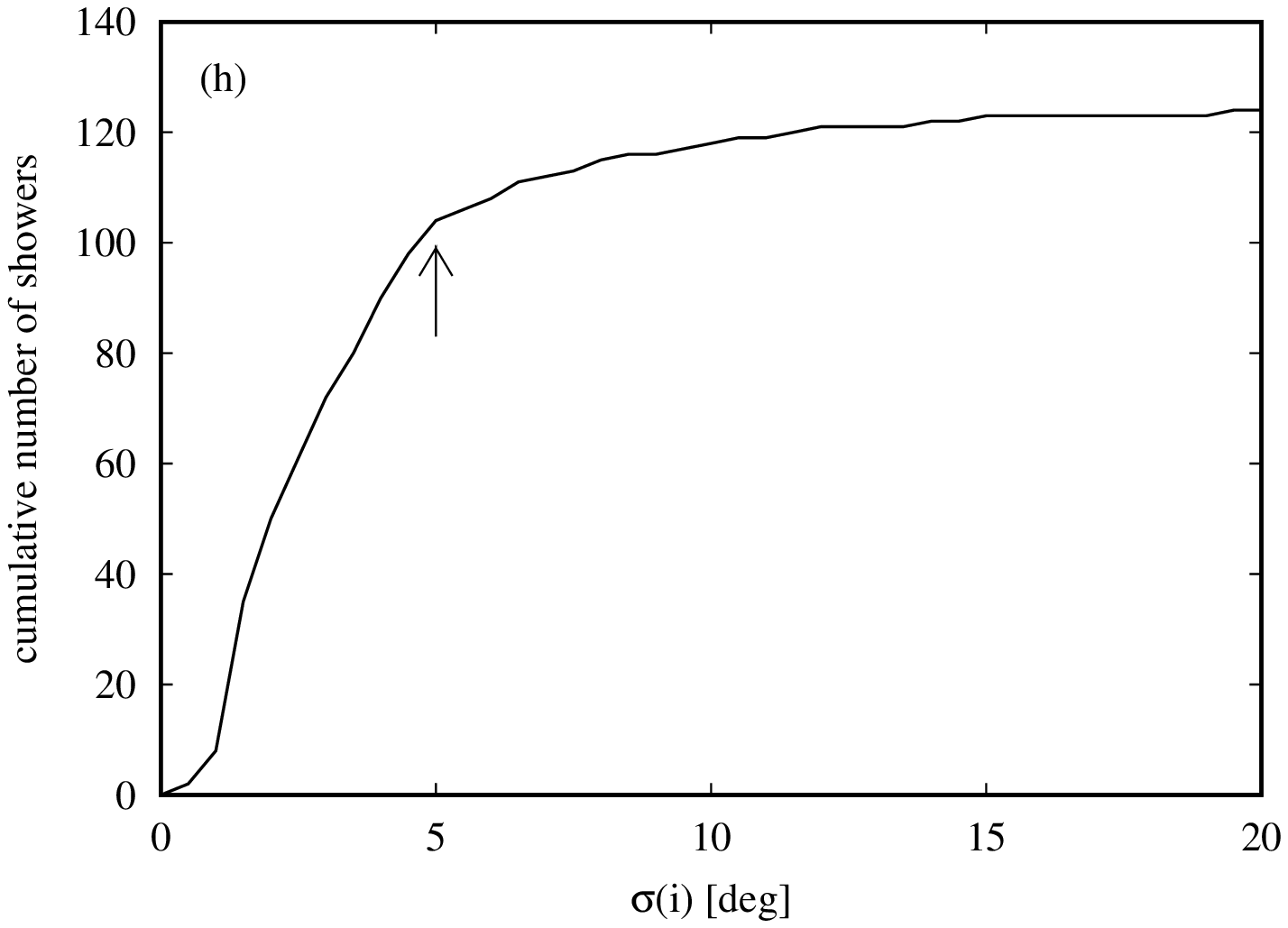}
               }
\caption{Cumulative distributions of the standard deviations of eight mean parameters of the showers in the IAU MDC with the Lookup Tables available.
}
\label{FIGdistrSIGs}
\end{figure*}
In Fig.~\ref{FIGdistrSIGs}, there are shown these distributions for eight shower's parameters. Specifically, panels (a), (b), (c), (f)--(h) of the figure show the distributions of $\sigma$ of the sun-centered ecliptic coordinates of mean radiant ($\lambda_{sc},\beta$), geocentric velocity ($V_g$), argument of perihelion ($\omega$), longitude of ascending node ($\Omega$), and inclination ($i$), respectively.
In these distributions, the steep increase is followed by a quasi constant behaviour (a plateau). It means that an essential part of the considered showers has the 1-$\sigma$ value of given mean parameter within the interval delimited by the beginning of the plateau.

In more detail we found $\sigma (\lambda_{sc}) = 5^{\circ}$, $\sigma (\beta) = 3^{\circ}$, $\sigma (V_{g}) = 2.5\,$km$\,$s$^{-1}$, $\sigma (\omega) = 9^{\circ}$, $\sigma (\Omega) = 7.5^{\circ}$, and $\sigma (i) = 5^{\circ}$. These values can be regarded as ``maximum $\sigma$s''. In Fig.~\ref{FIGdistrSIGs}, each maximum $\sigma$ is indicated with an arrow.
In the proposed method to find the duplicate solutions, the maximum $\sigma$, instead of $3\sigma$, is regarded as a critical difference of mean values of given parameter. If this difference is larger than maximum $\sigma$ at least for one parameter, then the examined solutions are autonomous or false-positive. Otherwise these are the duplicate solutions.

In the cumulative distributions of the 1-$\sigma$s of mean perihelion distance and eccentricity (Fig.~\ref{FIGdistrSIGs}d,e), there is no clear constant behaviour (maybe, we could consider $\sigma (q) = 0.08\,$au as the maximum deviation of $q$; but this is relatively large value). These two parameters are therefore useless for our purpose and are therefore not taken into account. 

 Finally, we take into account the 1-$\sigma$s of sun-centered ecliptic longitude, ecliptic latitude, geocentric velocity, argument of perihelion, longitude of ascending node, and inclination.
Hence, we propose that two showers A and B are considered duplicates, if the following conditions apply:
\begin{equation}
\begin{array}{ll}
\left |(\lambda_{sc})_A - (\lambda_{sc})_B \right |&< \, \sigma (\lambda_{sc}) = 5^{\circ}\\
\left |\beta_A-\beta_B\right |                     &<\, \sigma(\beta) = 3^{\circ}\\
\left |V_{gA}-V_{gB}\right |                       &<\, \sigma (V_g)  = 2.5\,{\rm km}\,{\rm s}^{-1}\\
\left |\omega_A - \omega_B\right | &<\, \sigma (\omega) = 9^{\circ}\\
\left |\Omega_A - \Omega_B \right |&<\, \sigma (\Omega) = 7.5^{\circ}\\
\left | i_A - i_B \right |&<\, \sigma (i) = 5^{\circ}\\
\end{array}
\label{warunkiMS}
\end{equation}
The maximum-$\sigma$ criterion (\ref{warunkiMS}) allows us to determine whether we are dealing with a duplicate or a false duplicate for a pair of solutions only. 
Therefore, in order to determine whether we are dealing with a group of duplicates of a given shower, it is necessary to apply criterion (\ref{warunkiMS}) in the cluster analysis.  Analogous to method I, we used a cluster analysis algorithm based on a single linking procedure.  In this approach, the inequalities (\ref{warunkiMS}) play an identical role to the D-function in cluster analysis among orbital data. 
 As in method-I, cluster analyses was performed separately in P1- 494 orbits and P2- 688 orbits. 
 \subsection{Method-I and method-II, important differences }
 \label{method-differences}
 The duplicate search methods used in this study are not equivalent to each other. 
 In our opinion, this is an advantageous property, as it strengthens the reliability of convergent results obtained by these methods.
 Both methods use the same cluster analysis algorithm, but the way of evaluating the similarity of the parameters of two meteor showers is fundamentally different. 
 
\begin{figure}
    \centering
    \includegraphics[width=0.45\textwidth, angle=0]{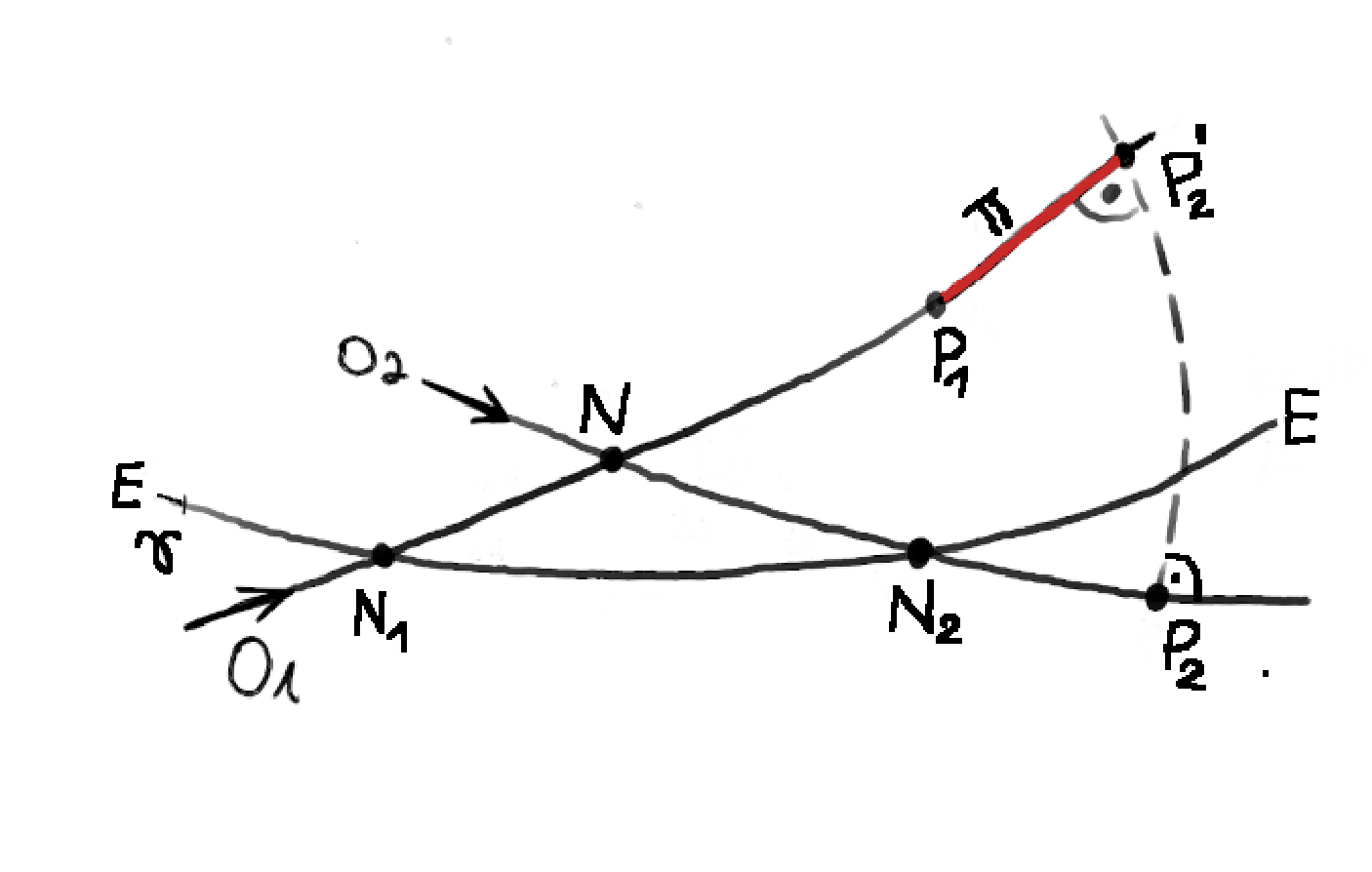}
\caption{Illustration of the difference in the orientation of the apsid lines in $D_H$-function. $E$ --- stands for ecliptic; $O_1$  denotes the orbit of the northern branch of the meteoroid stream; $O_2$ --- the orbit of the  southern branch of the meteoroid stream, $N_1, N_2$ are ascending and descending nodes of the orbits where meteors have been observed; $P_1, P_2$ denotes perihelion points of both orbit; $N$ --- common node of both orbit; $\pi$ --- in red, denotes the difference in the orientation of the apsid lines of the two orbits, measured from the common node $N$.}
\label{fig:ilustration_pi}
\end{figure}

 In the method-I, the heliocentric Keplerian elements of the orbits are compared. With the $D_H$ function used, differences in eccentricities, perihelion distances, and inclinations of the orbits are measured. However, the difference in the orientation of the apsidal lines is calculated as the difference in the angular positions of the perihelion points measured from the common point of intersection of the two orbits. Figure \ref{fig:ilustration_pi}
 illustrates this issue.  However, this means that in an automatic search by method-I, it will not always be possible to identify as separate groups the twin meteoroid streams or the northern and southern branches of the showers. Their separation will have to be done 'manually'. For example, we may encounter such situations in the case of the Orionids and eta-Aquariids showers, or in the case of the Northern and Southern Taurids.   

 For the method-II, this kind of situation will not occur. The similarity of the showers is evaluated separately for each parameter, geocentric and heliocentric. We will return to this issue in later sections of our paper. 
 %
\section{General results}
\label{results}
General results are collected in Table \ref{tab:summary}.  
\begin{table}
\caption{Total number of multi-solution showers $N_{M}$, number and percentage of duplicates $N_{D}$ off all MSS identified among $1182$ showers in the MDC database by method-I and method-II.
The last line lists the values of these figures based on the findings of the data providers to the MDC database.}
\footnotesize
\begin{center}
\begin{tabular}{l r r r }
\hline
\multicolumn{1}{c}{} & \multicolumn{1}{c}{$N_{M}$} & \multicolumn{1}{c}{$N_{D}$} & \multicolumn{1}{c}{$N_{D}$\%}  \\
\hline
\hline
Method I        &  85   & 456 & 38.7 \\
Method II       & 142  & 433 & 36.6 \\
In MDC          & 185  & 532 & 45.0 \\
\hline\hline
\end{tabular}
\end{center}
\label{tab:summary}
\normalsize
\end{table}
%

{\em Method-I}. Using $D_H$ function, among $1182$ orbits jointly  $85$ MSS was found comprising 456 duplicates. Some MSS consisted of only $2$ solutions, the largest consisted of $62$ solutions. The percentage of all duplicates was 38.7\% of the shower solutions taken from our $1182$ subset of the MDC. 

{\em Method-II}. Applying the maximum-sigma approach among $1182$ showers, total $142$ MSS were identified, comprising $433$ duplicates, they accounted for $36.6$\% of the MDC showers. 

For the purpose of comparison, the last row of Table~\ref{tab:summary} shows the corresponding values calculated based on the data provided to the MDC by their authors. As suggested by these authors, it was found that in our subset of the MDC, $532$ ($45.0$\%) duplicates represent $185$ multi-solution meteor showers.

This outcome allows us to conclude that method-I and method-II do not produce results fully  consistent with what is stated in the MDC. 
Compared to MDC, our methods produce smaller numbers of MSS groups ($85$and $142$ in comparison to $185$ stated in the MDC) which contain  $\sim100$ fewer duplicates (456, $433$ in comparison to $532$ according to the MDC). 
These are significant differences in the number of MSS as well as the overall number of duplicates, suggesting that authors providing meteor data to the MDC quite often misclassify their showers. We found that about dozens of the showers supplied to the MDC are unlikely to be duplicates.   

In Table \ref{tab:summary} we can clearly see that  method-I and method-II are not equivalent. 
This does not mean, however, that the methods used produce completely inadequate results. As far as the total number of duplicates is concerned, both methods I and II gave convergent results, 456 and $433$ duplicates, respectively.  The results obtained differ for the number of MSS identified, what means that the duplicates were classified into groups with different numbers of members, see Table \ref{tab:MSS_szczegoly}.
 Before presenting and discussing the results obtained, we would like to recall here, that for the method-I it was possible to estimate the probability of identifying MSS groups on a chance basis, it was less than $1$ per cent, see Section~\ref{Method-1}. For method-II, we do not have such an estimate. Similarly, we have no idea what is the reliability of the MSS findings made by the authors providing data to the MDC database. 

\begin{table} 
\caption{The table shows the numbers $N_{MSS}$ of multi-solution showers consisting of $N_{DU}$ duplicates, identified by method-I and method-II. The values given in the second column were obtained based on the classification of the showers given in the MDC.}
\footnotesize
\begin{center}
\begin{tabular}{c| c c c }
\multicolumn{1}{c|}{} & \multicolumn{3}{c}{$N_{MSS}$} \\
\multicolumn{1}{c|}{$N_{DU}$} & \multicolumn{1}{c}{MDC} & \multicolumn{1}{c}{M-I} & \multicolumn{1}{c}{M-II} \\
\hline
\hline
  2 & 106  & 27 &  81 \\
  3 &  44  & 19 &  30 \\
  4 &  13  & 10 &  11 \\
  5 &  12  &  9 &  12 \\
  6 &   4  &  4 &   3 \\
  7 &   2  &  1 &   1 \\
  8 &   1  &  2 &   0 \\
  9 &   1  &  1 &   1 \\
 10 &   1  &  2 &   1 \\
 11 &   1  &  3 &   0 \\
 12 &   0  &  2 &   0 \\
 13 &   0  &  1 &   0 \\
 14 &   0  &  0 &   0 \\
 15 &   0  &  2 &   0 \\
 16 &   0  &  0 &   1 \\
 17 &   0  &  0 &   1 \\
 18 &   0  &  0 &   0 \\
 ...&...   &...  &...  \\
 22 &   0  &  1 &   0 \\
 23 &   0  &  0  & 0 \\
 ...&...   &...  &...  \\
 62 &   0  &   1 &   0 \\
\hline
\hline
\end{tabular}
\end{center}
\label{tab:MSS_szczegoly}
\normalsize
\end{table}
\section{Results and discussion. Less complex cases.}  
\label{rezultatyszczegoly}
More detailed results are given in Table~\ref{tab:MSS_szczegoly}. In the Table, we see how many of the MSS were identified by method-I and method-II, and how many were reported in the MDC database. One can see that, in the case of a small number of duplicates ($N_{DU}>3$), the number of MSS reported in the MDC and obtained by method-II are far greater than the number of MSS identified by method-I. On the other hand, the most numerous MSS group containing $62$  duplicates were identified only by the method-I. 
At this stage, it is clear to us that the methods used in this study (and probably any other methods) cannot identify MSS groups as they are given in the MDC. Among other things because 
the authors providing data to the MDC followed unknown different criteria than we did in classifying the shower into a particular group. 
However, we are confident that the proposed approach of verifying the content of duplicates in the MDC database allows to eliminate all obvious problematic cases, thus improving the MDC database. 

In the following subsections \ref{dyskusja-1}--\ref{dyskusja-2}, we present results obtained directly (automatically) using our software. In subsection \ref{dyskusja-3}--\ref{dyskusja-4} we present results that, due to the nature of the method-I and also due to the limitations of statistical determination of orbital similarity thresholds --- required additional action on our part.    
\subsection{New MSS identified in the present study }
\label{dyskusja-1}
Among the $1182$ shower solutions taken from the MDC, we identified $7$ new MSS, listed in Table~\ref{tab:newMSS}. Two duplicates were identified in each new MSS. The reliability of this result is confirmed by the low threshold values of orbital similarity, corresponding to $<1$\% probability of identifying each pair by chance, by method-I, and by the fact that they were also identified by method-II.
\begin{table} 
\begin{center}
\footnotesize
\caption{List of $7$ new MSS found among $1182$ meteor data identified by both of the applied methods. The individual columns means: the MDC codes designation of the MSS, $N$ --- number of duplicates provided for each MSS by both methods, $DH_{M_I}$  --- for method-I, the maximum acceptable threshold value to reliably identify each pair. }
\begin{tabular}{r l c c }
\multicolumn{1}{c}{ }&\multicolumn{1}{c}{ Shower Code}& \multicolumn{1}{c}{$N$} &  \multicolumn{1}{c}{$DH_{M_I}$}\\
\hline
\hline
  1 &   0246/00/AMO &      2 &  0.059  \\ 
    &   1196/00/ZCM &        &          \\
  2 &   0575/00/SAU &      2 &  0.059  \\
    &   1151/00/NPA &        &           \\
  3 &   0658/00/EDR &      2 &  0.039  \\ 
    &   1110/00/CEP &        &           \\
  4 &   1071/00/IHD &      2 &  0.059  \\ 
    &   1108/00/IHR &        &          \\
  5 &   1089/00/CTS &      2 &  0.039  \\
    &   1157/00/FCD &        &          \\
  6 &   1097/00/DOH &      2 &  0.059  \\
    &   1161/00/THT &        &          \\
  7 &   1099/00/JED &      2 &  0.059  \\ 
    &   1107/00/JID &        &           \\
\hline
\hline
\end{tabular}
\label{tab:newMSS}
\end{center}
\normalsize
\end{table}
\subsection{MSS fully confirmed in the present search}
\label{dyskusja-2}
Table~\ref{tab:exactlyMSS} lists the codes of 30 MSS identified identically by the two methods used. In each of the MSS, the same duplicates and the same number of duplicates were identified by both methods, exactly as they are listed in the MDC database. 
\begin{table}
\begin{center}
\footnotesize
\caption{List of 30 MSS for which exactly the same duplicates were found by both of the applied methods and as given in the MDC. The individual columns means: the MDC codes designation of the MSS, $N_{MDC}$ --- number of duplicates provided for the MSS in the MDC, $DH_{Min}$ --- minimum threshold required to identify all members of the MSS's provided in the MDC. The last column $DH_{M_I}$ gives the maximum acceptable threshold value to reliably identify, using Method-I, a group of the size in column $N_{MDC}$.}
\begin{tabular}{r l c c c}
\multicolumn{1}{c}{ }&\multicolumn{1}{c}{ Shower Code}& \multicolumn{1}{c}{$N_{MDC}$} & \multicolumn{1}{c}{$DH_{Min}$} &  \multicolumn{1}{c}{$DH_{M_I}$}\\
\hline
\hline
  1  & 0004/00/GEM &    5 &  0.020 &  0.039 \\ 
  2  & 0006/00/LYR &    6 &  0.045 &  0.059 \\ 
  3  & 0007/00/PER &    3 &  0.011 &  0.059 \\ 
  4  & 0010/00/QUA &    6 &  0.050 &  0.059 \\ 
  5  & 0016/00/HYD &    3 &  0.095 &  0.136 \\ 
  6  & 0019/00/MON &    5 &  0.044 &  0.079 \\ 
  7  & 0022/00/LMI &    2 &  0.037 &  0.059 \\ 
  8  & 0110/00/AAN &    5 &  0.109 &  0.136 \\ 
  9  & 0171/00/ARI &    5 &  0.090 &  0.099 \\ 
 10  & 0184/02/GDR &    2 &  0.006 &  0.059 \\ 
 11  & 0191/00/ERI &    2 &  0.049 &  0.059 \\ 
 12  & 0250/00/NOO &    3 &  0.053 &  0.079 \\ 
 13  & 0323/00/XCB &    4 &  0.100 &  0.136 \\ 
 14  & 0331/00/AHY &    3 &  0.050 &  0.059 \\ 
 15  & 0341/02/XUM &    2 &  0.025 &  0.059 \\ 
 16  & 0429/00/ACB &    3 &  0.041 &  0.059 \\ 
 17  & 0444/00/ZCS &    2 &  0.045 &  0.059 \\ 
 18  & 0446/00/DPC &    2 &  0.010 &  0.039 \\ 
 19  & 0458/00/JEC &    3 &  0.081 &  0.136 \\ 
 20  & 0464/00/KLY &    3 &  0.030 &  0.039 \\ 
 21  & 0465/00/AXC &    3 &  0.064 &  0.136 \\ 
 22  & 0497/01/DAB &    2 &  0.048 &  0.059 \\ 
 23  & 0506/00/FEV &    2 &  0.055 &  0.059 \\ 
 24  & 0519/00/BAQ &    2 &  0.041 &  0.059 \\ 
 25  & 0523/00/AGC &    2 &  0.030 &  0.059 \\ 
 26  & 0525/00/ICY &    2 &  0.031 &  0.039 \\ 
 27  & 0526/00/SLD &    3 &  0.032 &  0.059 \\ 
 28  & 0529/00/EHY &    3 &  0.041 &  0.059 \\ 
 29  & 0570/00/FBH &    2 &  0.024 &  0.059 \\ 
 30  & 1044/00/EPU &    2 &  0.034 &  0.059 \\ 
\hline
\hline
\end{tabular}
\label{tab:exactlyMSS}
\end{center}
\normalsize
\end{table}
As one can see in Table~\ref{tab:exactlyMSS} the $DH_{M_I}$ threshold values with which the method-I identified MSS are always greater than the corresponding $DH_{min}$ values. This supports the statement that among these MSS we are not dealing with false duplicates. The authors providing the data for these showers classified them in the right way. 
At the same time, this means that among the remaining 150 cases, some misclassification of duplicates is possible. 
\subsection{MSS not identified in the present search}
\label{dyskusja-3}
Among the $185$ of MSS reported in the MDC database, 56 groups were not identified as MSS by the two methods used.

The Table~\ref{tab:falseMSS_1} states the orbital $DH$- similarity values of the showers associated with that MSS. The $DH_{min}$ is the smallest threshold that would have to be used in the cluster analysis to identify by method-I all members of the given  MSS as reported  in the MDC database. The $DH_{M_I}$ is the maximum threshold with which the given MSS should be identified by the method-I. It should be noted that, as mentioned in section~\ref{Method-1}, for P-1- and P2- partitions the orbital similarity threshold values were determined separately. 

\begin{table}
\begin{center}
\footnotesize
\caption{56 MSS listed in the MDC for which no duplicates were found by any of the applied methods. The contents of the table's columns are as in Table-\ref{tab:exactlyMSS}. $N_r$ gives the number of solutions remaining in a given MSS after our verification.
$N_r=1$ means that the group consists of only one solution, so it also has no MSS status.
In italic, particularly questionable cases of MSS are highlighted in column $DH_{Min}$.}

\begin{tabular}{r l c c c c}
\multicolumn{1}{c}{ }&\multicolumn{1}{c}{ Shower Code}& \multicolumn{1}{c}{$N_{MDC}$} & \multicolumn{1}{c}{$DH_{Min}$} &  \multicolumn{1}{c}{$DH_{M_I}$} &  \multicolumn{1}{c}{$N_r$}\\
\hline
\hline
  1  & 0003/00/SIA &    2 &  0.065 &  0.039 & 0\\ 
  2  & 0011/00/EVI &    3 &  \bo{0.322} &  0.079 & 1\\ 
  3  & 0025/00/NOA &    2 &  \bo{0.210} &  0.039 & 1\\ 
  4  & 0032/00/DLM &    2 &  \bo{0.321} &  0.059 & 1\\ 
  5  & 0040/00/ZCY &    2 &  \bo{0.259} &  0.059 & 0\\ 
  6  & 0076/00/KAQ &    2 &  \bo{0.142} &  0.039 & 0\\ 
  7  & 0088/00/ODR &    2 &  \bo{0.362} &  0.059 & 0\\ 
  8  & 0093/00/VEL &    2 &  \bo{0.412} &  0.059 & 0\\ 
  9  & 0100/00/XSA &    2 &  \bo{0.209} &  0.039 & 0\\ 
 10  & 0105/00/OCN &    2 &  \bo{0.134} &  0.059 & 0\\ 
 11  & 0106/00/API &    2 &  \bo{0.329} &  0.059 & 0\\ 
 12  & 0107/00/DCH &    3 &  \bo{0.358} &  0.136 & 0\\ 
 13  & 0108/00/BTU &    2 &  \bo{0.344} &  0.059 & 0\\ 
 14  & 0113/00/SDL &    2 &  \bo{0.172} &  0.039 & 0\\ 
 15  & 0118/00/GNO &    2 &  \bo{0.798} &  0.059 & 0\\ 
 16  & 0121/00/NHY &    3 &  \bo{0.269} &  0.079 & 0\\ 
 17  & 0124/00/SVI &    2 &  \bo{0.226} &  0.039 & 0\\ 
 18  & 0127/00/MCA &    2 &  0.041 &  0.039 & 0\\ 
 19  & 0128/00/MKA &    2 &  \bo{0.315} &  0.039 & 0\\ 
 20  & 0133/00/PUM &    2 &  \bo{0.139} &  0.039 & 0\\ 
 21  & 0150/01/SOP &    2 &  \bo{0.343} &  0.039 & 0\\ 
 22  & 0151/00/EAU &    2 &  \bo{0.155} &  0.059 & 0\\ 
 23  & 0152/00/NOC &    4 &  \bo{0.644} &  0.182 & 0\\ 
 24  & 0154/00/DEA &    2 &  \bo{0.178} &  0.039 & 1\\ 
 25  & 0167/00/NSS &    2 &  \bo{0.341} &  0.039 & 0\\ 
 26  & 0170/00/JBO &    2 &  \bo{0.220} &  0.039 & 1\\ 
 27  & 0179/00/SCA &    2 &  \bo{0.234} &  0.039 & 0\\ 
 28  & 0183/00/PAU &    3 &  \bo{0.294} &  0.136 & 0\\ 
 29  & 0186/00/EUM &    2 &  \bo{0.194} &  0.039 & 1\\ 
 30  & 0188/00/XRI &    3 &  \bo{0.534} &  0.079 & 0\\ 
 31  & 0189/00/DMC &    2 &  \bo{0.454} &  0.039 & 0\\ 
 32  & 0197/00/AUD &    2 &  \bo{0.316} &  0.039 & 1\\ 
 33  & 0199/00/ADC &    2 &  \bo{0.107} &  0.039 & 0\\ 
 34  & 0202/00/ZCA &    2 &  \bo{0.434} &  0.039 & 0\\ 
 35  & 0219/00/SAR &    4 &  \bo{0.373} &  0.099 & 0\\ 
 36  & 0220/00/NDR &    2 &  \bo{0.144} &  0.039 & 0\\ 
 37  & 0233/00/OCC &    2 &  \bo{0.160} &  0.039 & 0\\ 
 38  & 0253/00/CMI &    2 &  \bo{0.347} &  0.039 & 0\\ 
 39  & 0254/00/PHO &    2 &  \bo{0.199} &  0.039 & 1\\ 
 40  & 0321/00/TCB &    2 &  0.089 &   0.059 & 0\\ 
 41  & 0324/00/EPR &    2 &  \bo{0.250} &  0.059 & 0\\ 
 42  & 0326/00/EPG &    2 &  \bo{0.175} &  0.059 & 0\\ 
 43  & 0327/00/BEQ &    2 &  \bo{0.377} &  0.059 & 0\\ 
 44  & 0334/01/DAD &    2 &  0.110 &   0.059 & 1\\ 
 45  & 0347/00/BPG &    2 &  \bo{0.230} &  0.059 & 0\\ 
 46  & 0372/00/PPS &    2 &  \bo{0.350} &  0.059 & 1\\ 
 47  & 0386/00/OBC &    2 &  \bo{0.266} &  0.059 & 0\\ 
 48  & 0392/00/NID &    2 &  \bo{0.317} &  0.059 & 1\\ 
 49  & 0490/00/DGE &    2 &  \bo{0.149} &  0.039 & 1\\ 
 50  & 0507/00/UAN &    2 &  \bo{0.326} &  0.059 & 0\\ 
 51  & 0512/00/RPU &    2 &  \bo{0.370} &  0.059 & 0\\ 
 52  & 0555/00/OCP &    2 &  \bo{0.282} &  0.059 & 0\\ 
 53  & 0574/00/GMA &    2 &  \bo{0.480} &  0.059 & 0\\ 
 54  & 0644/00/JLL &    2 &  \bo{0.256} &  0.039 & 1\\ 
 55  & 0709/00/LCM &    2 &  \bo{0.142} &  0.039 & 0\\ 
 56  & 1048/00/JAS &    2 &  \bo{0.150} &  0.059 & 0\\ 
\hline
\hline
\end{tabular}
\label{tab:falseMSS_1}
\end{center}
\normalsize
\end{table}

In Table~\ref{tab:falseMSS_1}, almost all the threshold values $DH_{Min}$ are too large (the values highlighted in bold) compared to the limiting $DH_{M_I}$ values, the maximum acceptable threshold value to reliably identify, using method-I, a group of of the size in column $N_{MDC}$. For example, $DH_{min}$=$0.315$ for 128/00/MKA or $DH_{min}$=$0.798$ for 0118/00/GNO are much to large, because to identify a MSS with 2 duplicates only, the acceptable threshold value $DH_{M_I}=0.040$ or $DH_{M_I}=0.063$. Such small thresholds were used in method-I in the cluster analysis and it ensures that the probability of identifying a group of two duplicates by chance was less than $1$\%.

For 40 MSS, all duplicates were rejected, and the rejected showers did not enter another MSS group. In $13$ MSS, on the other hand, after rejecting false-duplicates, only single solutions remained, which often entered another MSS group.

As a result, we can claim that except of 3/00/SIA, 
127/00/MCA, 321/00/TCB remaining 53 MSS listed in Table~\ref{tab:falseMSS_1}  should be considered as the false MSS, and duplicates assigned with them should receive a stand-alone shower status in the MDC database. 

We also recognise the possibility that, in the MDC database, there are erroneous values among the shower parameters, the codes for which are given in Table~\ref{tab:falseMSS_1}. For this reason, it would be appropriate to ask  (if possible) the authors providing the data to the MDC to check the correctness of the shower parameter values or the MSS classifications  accordingly. 

As an example of the problematic case, we will present the details of the shower 152/NOC  (Table~\ref{tab:falseMSS_1}, line 21) with very high $DH_{min}=0.644$ value. In the MDC database the 152/NOC shower, named the Northern Daytime omega-Cetids,  is represented by 4 solutions listed in Table~\ref{tab:NOC}.
\begin{table}
\begin{center}
\footnotesize
\caption{List of four solutions of averaged Keplerian orbital elements of the 152/NOC shower, the  Northern Daytime omega-Cetids. The $N$ column lists the number of averaged orbits of each solution. The data were taken from the MDC database, status December 2022, see \citet{2017P&SS..143....3J, 2023A&A...671A.155H}.}
\begin{tabular}{r c c c c c c }
\multicolumn{1}{c}{ }&\multicolumn{1}{c}{q [au]} & \multicolumn{1}{c}{e} &  \multicolumn{1}{c}{$\omega$}[deg] &  \multicolumn{1}{c}{$\Omega$[deg]} &  \multicolumn{1}{c}{inc [deg]} &  \multicolumn{1}{c}{N} \\
\hline
\hline
1  & 0.108   & 0.889   &  25.61  &  47.79  &  42.00  & 16  \\
2  & 0.17    & 0.93    &  42.61  &  64.36  &  10.20  & 10  \\
3   & 0.118   & 0.9256  & 33.1    & 45.1    & 34.2    & 1256 \\
4   & 0.096   & 0.956   & 31.4    & 53.6    & 40.2    & 12 \\
\hline
\hline
\end{tabular}
\label{tab:NOC}
\end{center}
\normalsize
\end{table}
As can be seen in Table~\ref{tab:NOC}, compared to the others, solution 2 has clearly different values of orbital angular elements. In particular, it has a much smaller orbit inclination value. 
Hence, in our opinion, solution 2 was wrongly classified for this MSS shower. This  solution was published by \citet{1964AuJPh..17..205N}, the shower was identified among radio-observed meteors. Nilsson did not propose a name for this shower, and in his paper in Table 4 he identified it as 'Gr. 61.5.3'. This shower consists of 10 members. Nilsson's identification turned out to be correct, it was confirmed by a completely different cluster analysis method by \citet{1999md98.conf..307J}; in Table 2 of this work, the stream was named mu-Arietids. And it is unclear to us why this shower was classified in \citet{2006mspc.book.....J} book as another solution of the 152/NOC shower. Additional cluster analysis performed without this solution showed that the other remaining 3 solutions of the 152/NOC were identified exactly as given in the MDC database.

The example discussed above leads us to believe that we have a similar situation for many of the other showers listed in Table~\ref{tab:NOC} as well.  These will be presented and discussed in details in our next publication.
\subsection{MSS identified only by method-II}
%
The results presented in the previous sections were for MSS identically identified or unidentified by both method I and method II. 
In the current chapter, we present results obtained only with method-II. In Table~\ref{tab:confirmedbyMII}, we report 20 MSS showers occurring in the MDC database and fully confirmed but by method-II only. At the same time, according to method-I, none of the duplicates in question are members of any MSS.    
\begin{table}
\begin{center}
\footnotesize
\caption{
List of 20 MSS for which exactly the same duplicates --- as given in the MDC ---  were found only by method-II. 
The contents of the table's columns are as in Table-\ref{tab:exactlyMSS}.}
\begin{tabular}{r l c c c}
\multicolumn{1}{c}{ }&\multicolumn{1}{c}{ Shower Code}& \multicolumn{1}{c}{$N_{MDC}$} & \multicolumn{1}{c}{$DH_{Min}$} &  \multicolumn{1}{c}{$DH_{M_I}$}\\
\hline
\hline
  1  & 0112/01/NDL &    2 &  0.070 &  0.039 \\
  2  & 0153/00/OCE &    3 &  0.081 &  0.079 \\
  3  & 0322/00/LBO &    2 &  0.072 &  0.059 \\
  4  & 0333/00/OCU &    2 &  0.100 &  0.059 \\
  5  & 0348/00/ARC &    2 &  0.096 &  0.059 \\
  6  & 0388/00/CTA &    3 &  0.127 &  0.079 \\
  7  & 0390/00/THA &    2 &  0.135 &  0.039 \\
  8  & 0431/00/JIP &    2 &  0.062 &  0.059 \\
  9  & 0439/00/ASX &    2 &  0.077 &  0.059 \\
 10  & 0450/00/AED &    2 &  0.061 &  0.059 \\
 11  & 0486/00/NZP &    2 &  0.041 &  0.039 \\
 12  & 0510/00/JRC &    2 &  0.068 &  0.059 \\
 13  & 0537/00/KAU &    2 &  0.141 &  0.059 \\
 14  & 0538/00/FFA &    2 &  0.092 &  0.039 \\
 15  & 0549/00/FAN &    2 &  0.105 &  0.059 \\
 16  & 0558/00/TSM &    2 &  0.076 &  0.059 \\
 17  & 0569/00/OHY &    2 &  0.082 &  0.059 \\
 18  & 0651/00/OAV &    2 &  0.096 &  0.039 \\
 19  & 0746/00/EVE &    2 &  0.085 &  0.059 \\
 20  & 1106/00/GAD &    2 &  0.069 &  0.059 \\
\hline
\hline
\end{tabular}
\label{tab:confirmedbyMII}
\end{center}
\normalsize
\end{table}
The reason of it is that --- according to MDC --- the duplicates belonging to these showers are too dissimilar. Indeed, the acceptable $DH_{MI}$ thresholds given in Table~\ref{tab:confirmedbyMII} are always less than those that would have to be used in a cluster analysis to consider duplicates belonging to a given MSS as properly classified. However, unlike the contents of Table~\ref{tab:falseMSS_1}, this time the differences between $DH_{min}$ and $DH_{MI}$ values are noticeably smaller.%

There could be two reasons for this, either the thresholds used in method-I are too stringent (which is possible), or method-II is too tolerant; e.g. because method-II does not take into account differences in the orbital elements e and q, which determine the shape and size of the orbit.  
It is almost certain that introducing these elements into the conditions \ref{warunkiMS} would at least, reduce the number of MSS listed in Table~\ref{tab:confirmedbyMII}. 

But other reasons are also possible, such as errors in the parameters of the showers supplied to the MDC base and also failure to meet the assumptions we mentioned in section~\ref{methodology}, namely the impact of individual averaging of meteoroid parameters causing additional differences between the parameters to increase.

Due to the small differences between the $DH_{min}$ and $DH_{MI}$ values, at this stage of our research, we considered the MSS showers confirmed by either method to be fully confirmed.
\section{Results and discussion. More difficult cases.}
\label{alsoconfirmed}
The inconsistencies in the results obtained by the two methods, discussed in the previous section do not exhaust all the inconsistencies we encountered in our study. We will present them in the following subsections.
\subsection{MSS sufficiently confirmed}
\label{dyskusja-4}
The results given in the previous sections were selected 'automatically' using our software.  
Here we present results confirming the existence of a given MSS in the MDC database. However, some of them were not confirmed 'automatically' by both methods. Sometimes the members of the given MSS were picked out of the more complex identified groups in a 'manual' manner. 

Table~\ref{tab:sufficientlyMSS} lists seven MSS, identified 'automatically',  for which the numbers of duplicates identified by both methods are more or less close to, the number as reported in the MDC database.
Compared to what is given in the MDC database, the differences are quantitative, consisted of deleting or adding new members to the group. Hence, in our opinion, all MSS listed in Table~\ref{tab:sufficientlyMSS} can be considered sufficiently confirmed.

\begin{table}
\begin{center}
\scriptsize
\footnotesize
\caption{List of seven MSS for which the number of the same duplicates found by method-I and method-II is close to the corresponding number in the MDC database. The individual columns means: the MDC codes designation of the MSS, $N$ --- number of duplicates provided for the MSS in the MDC, $N_{{I}}$, $N_{{II}}$ ---number of duplicates found by method-I and method-II;  $D_{Min}$ --- minimum threshold required to identify all members of the MSS's provided in the MDC; $D_{M_{I}}$ gives the maximum acceptable threshold value to reliably identify by method-I a group of of the size in column $N$.}
\begin{tabular}{r l c c c c c }
\multicolumn{1}{c}{ }&\multicolumn{1}{c}{ Shower Code}& \multicolumn{1}{c}{$N$} & 
\multicolumn{1}{c}{$N_{I}$} & \multicolumn{1}{c}{$N_{{II}}$} &  
\multicolumn{1}{c}{$D_{Min}$} &  \multicolumn{1}{c}{$D_{M_I}$} \\
\hline
\hline
  1  & 0001/00/CAP &    9 &   13 &   10 & 0.062 &   0.079  \\ 
  2  & 0005/00/SDA &    8 &    8 &    9 & 0.078 &  0.079  \\ 
  3  & 0009/00/DRA &    4 &    3 &    3 & 0.186 &  0.079  \\ 
  4  & 0015/03/URS &    3 &    4 &    3 & 0.149 &  0.182  \\ 
  5  & 0319/00/JLE &    3 &    4 &    2 & 0.143 &  0.182  \\
  6  & 0502/00/DRV &    3 &    4 &    4 & 0.042 &  0.136  \\ 
  7  & 0505/00/AIC &    3 &    2 &    3 & 0.117 &  0.039  \\ 
\hline
\hline
\end{tabular}
\label{tab:sufficientlyMSS}
\end{center}
\normalsize
\end{table}
\subsection{MSS confirmed after manual action}
\label{dyskusja-5}
However, in a dozen cases, MSS identification required manual action in the results obtained automatically. Here we are referring to MSS identified with a distinctly larger number of duplicates, that included other groups treated as separate MSS in the MDC database. An extreme case was 2/000/STA (Southern Taurids) in which method-I identified $62$ duplicates.

The actual meteoroid streams may have a different structure in the parameter space used in the method-I and method-II. These parameters may occupy volumes that resemble hyper-spheres in five-dimensional space (Geminids), but equally well they may occupy more extended volumes (Taurids). 
The reason for the complexity of the results discussed in this section is related, among others, with the property of the single linking algorithm used in the cluster analysis in both methods. 
With this algorithm, the so-called chain effect is possible.  As a result of this, successively connected group members form a chain-like structure, which, if too high a threshold value is applied, leads to the identification of an unrealistic group of orbits. For these reasons, determining by statistical methods the correct threshold values of orbital similarity taking into account only the abundance of the meteoroid stream may lead to unrealistic results.

This chain effect is clearly stronger in the case of the method-I for the shower 2/00/STA --- Southern Taurids. An ecliptic group of 62 duplicates was identified by method-I, including all the duplicates of the Northern and Southern Taurids showers, but also 41 duplicates belonging to a dozen other MSS. 
Another reason of the variation in results is attributable to the properties of the parameter's similarity metrics used, as we mentioned in Section~\ref{method-differences}. 

Table ~\ref{tab:manuallyMSS} contains a list of 27 MSS obtained manually from results identified automatically. All of the MSS in this table can be considered confirmed, but we believe it would be advisable to examine each of them in a separate detailed study.  
\begin{table}
\begin{center}
\scriptsize
\footnotesize
\caption{List of 27 MSS for which, after manual action the number of the same duplicates found by method-I and method-II is close to the corresponding number in the MDC database. The contents of the column are as shown in Table~\ref{tab:sufficientlyMSS}. }
\begin{tabular}{r l c c c c c}
\multicolumn{1}{c}{ }&\multicolumn{1}{c}{ Shower Code}& \multicolumn{1}{c}{$N$} &  
\multicolumn{1}{c}{$D_{Min}$} & 
\multicolumn{1}{c}{$N_{I}$} &  
\multicolumn{1}{c}{$D_{M_I}$} & \multicolumn{1}{c}{$N_{{II}}$} \\
\hline
\hline
  1  & 0002/00/STA &   11 &  0.182 &  11     &  0.134  &    9  \\ 
  2  & 0008/00/ORI &    5 &  0.062 &   5     &  0.267  &    5  \\ 
  3  & 0012/00/KCG &    4 &  0.117 &   4     &  0.139  &    2  \\ 
  4  & 0013/00/LEO &    7 &  0.158 &   7     &  0.257  &    7  \\ 
  5  & 0017/00/NTA &   10 &  0.240 &  10     &  0.134  &   10  \\
  6  & 0021/00/AVB &    3 &  0.124 &  3 &  0.139  &    2   \\ 
  7  & 0025/01/NOA &    2 &  0.113 &   1     &  0.134  &   1   \\ 
  8  & 0028/00/SOA &    2 &  0.113 &   2     &  0.134  &   1   \\ 
  9  & 0031/00/ETA &    5 &  0.109 &   5     &  0.267  &   4  \\ 
 10  & 0033/00/NIA &    5 &  0.158 &   5     &      0.142    &   2 \\
 11  & 0096/00/NCC &    5 &  0.162 &   5     &  0.134  &    5  \\
 12  & 0097/00/SCC &    4 &  0.218 &   2     &  0.134  &    2 \\
 13  & 0156/00/SMA &    3 &  0.362 &   2     &    0.142      &    2\\
 14  & 0172/00/ZPE &    4 &  0.075 &   4     &  0.134  &   4 \\ 
 15  & 0215/00/NPI &    4 &  0.170 &   4     &  0.142  &   5 \\ 
 16  & 0216/00/SPI &    5 &  0.460 &   4     &  0.134  & 3 \\ 
 17  & 0256/00/ORN &    4 &  0.201 &   2     &  0.134  &    2 \\
 18  & 0257/00/ORS &    5 &  0.123 &   5     &  0.134  & 4 \\ 
 19  & 0343/02/HVI &    6 &  0.103 &   6     &  0.139  &  6   \\
 20  & 0428/00/DSV &    2 &  0.067 &   2     &  0.274 &    3 \\ 
 21  & 0456/00/MPS &    4 &  0.041 &   4     &  0.142 &    4 \\ 
 22  & 0460/00/LOP &    3 &  0.044 &   3     &  0.142 &    3 \\ 
 23  & 0466/01/AOC &    2 &  0.086 &   5     &  0.136 &    5 \\ 
 24  & 0 479/00/SOO &   3 &  0.072 &   3     &  0.257 &    3 \\
 25  & 0480/00/TCA &    3 &  0.066 &   3     &  0.257 &    4 \\ 
 26  & 0481/00/OML &    3 &  0.031 &   3     &  0.257 &    3 \\  
 27  & 0533/00/JXA &    3 &  0.037 &   3     &  0.226 &    4 \\ 
\hline
\hline
\end{tabular}
\label{tab:manuallyMSS}
\end{center}
\normalsize
\end{table}
%
\section{Conclusion and discussion of future action}
Two methods were used to search for MSS among $1182$ meteor shower solutions selected from MDC database.
The obtained results confirmed the effectiveness of the proposed approach of identifying duplicates. We have shown that in order to detect and verify duplicate meteor showers, it is possible to apply the objective proposal
instead of the subjective approach used so far. 

However, it appears that to reveal the duplicate showers is not a simple task. One must face to a wide variety of the properties of various meteor showers. There are known not only the compact showers with the parameters each ranging in a relatively narrow interval, but also the showers with a largely enough dispersed parameters. Sometimes, a shower possesses some structural features and it is only a matter of convention to regard these features as the substructures of given shower or as the autonomous showers.

In our study a number of results of varying significance were obtained: (i) seven new MSS represented by two or more parameter sets were discovered, (ii) for 30 MSS  there was full agreement between our results and those reported in the MDC
database, (iii) for 20 MSS the same duplicates as given in the MDC, were found only by one method, (iv)  we found 34 MSS for which the number of the same duplicates found by both method is close to the corresponding
number in the MDC database, (v) for 56 MSS listed in the MDC no duplicates were found by any of the applied methods. 

We consider the identification of 34 + 56 problematic cases in the MDC database, among which at least some duplicates were misclassified, to be a particularly important result. The correction of these cases will significantly improve the content of the MDC database.
As shown in Section \ref{dyskusja-3}, such an adjustment is possible, but it always requires a meticulous approach, so we decided to pursue it in subsequent studies.

Determining the correct duplicates of the MSS is important when it comes to giving a shower its established status and, consequently, its official shower name. In \citet{2023A&A...671A.155H} paper, one of the criteria that must be met for this purpose is --- the shower must be represented by at least two sets of parameters (duplicates) determined by independent authors.

Anyway, we are convinced that our work helps to identify the problems related to the duplicity problem and a wide discussion in the meteor-research community will follow. 
\begin{acknowledgements}
This work was supported, in part, by the VEGA - Slovak Grant Agency
for Science, grant No. 2/0009/22. 
This research has made use of NASA's Astrophysics Data System Bibliographic Services.
\end{acknowledgements}

%
%
\bibliographystyle{aa}
\bibliography{showerduplicates} 
\end{document}